# Tsallis Non-Extensive Statistics. Theory and Applications

G.P. Pavlos[1], L.P. Karakatsanis[1], M.N. Xenakis[2,] A.E.G. Pavlos [3]A.C. Iliopoulos [1], D.V. Sarafopoulos [1]

*[1] Department of Electrical and Computer Engineering, Democritus University of Thrace, 67100 Xanthi, Greece*
*[2] German Research School for Simulation Sciences, Aachen, Germany*
*[3] Department of Physics, Aristotle University of Thessaloniki, 54624 Thessaloniki, Greece*
*Email:* gpavlos@ee.duth.gr

**Abstract**

*In this study the q-statistics of Tsallis theory is testified in various complex physical systems. Especially the Tsallis q-triplet is estimated for space plasmas atmospheric dynamics and seismogenesis as well as for the brain and cardiac activity. The coincidence of theoretical predictions following after Tsallis theory and experimental estimations were found to be excellent in all cases*

**Keywords:** *Tsallis non-extensive statistics, Fractal Dynamics, Space Plasmas, Fractal Acceleration, Fractal Dissipation, Intermittent Turbulence.*

## 1. Introduction

Near thermodynamic equilibrium the statistics and dynamics are two separated but fundamental elements of the physical theory. Also at thermodynamical equilibrium nature reveals itself as a Gaussian and uncorrelated process simultaneously with unavoidable or inevitable and objective deterministic character. However modern evolution of the scientific knowledge reveals for the equilibrium characteristics of the physical theory as an approximation or the limit of more synthetic physical theory which is characterized as complexity. The new physical characteristics of complexity theory can be manifested as the physical system is removed far from equilibrium.

Far from equilibrium statistic and dynamics can be unified through the Tsallis nonextensive statistics included in Tsallis q- entropy theory [1] and the fractal generalization of dynamics included in theories developed by Ord [2] El-Naschies [3], Nottale [4], Castro [5], Zaslavsky [6], Shlesinger [7], Kroger [8] Tarassov [9], El-Nabulsi [10], Cresson [11], Coldfain [12] and others.

The new character of non-equilibrium physical theory has been verified by Pavlos [13] and Karakatsanis [14] until now in many cases especially in space plasmas and at other regions of natural systems [1].

The traditional scientific point of view is the priority of dynamics over statistics. That is dynamics creates statistics. However for complex system their holistic behaviour does not permit easily such a simplification and division of dynamics and statistics. Tsallis $q$ – statistics and fractal or strange kinetics are two faces of the same complex and holistic (non-reductionist) reality. As Tsallis statistics is an extension of B-G statistics, we can support that the thermic and the dynamical character of a complex system is the manifestation of the same physical process which creates extremized thermic states (extremization of Tsallis entropy), as well as dynamically ordered states. From this point of view the Feynman path integral formulation of physical theory [15] indicates the indivisible thermic and dynamical character of physical reality. After the Heisenberg's revolutionary substitution of physical magnitudes by operators and the probabilistic interpretation of Heisenberg operationalistic concepts and Schrodinger wave functions by Max Born [16] probability obtained clearly the physical character and physical reality, foundation of physics [17].

Today the probabilistic character of dynamics was extended out of quantum theory and obtained a deep meaning after the fractal extension of dynamics at all levels of physical microscopic or macroscopic reality [3-5, 9, 10, 12, 18], as well as the q-extension of statistics in Tsallis non-extensive entropy theory.

Prigogine [19] and Nicolis [20] were the principal leaders of an outstanding transition to the new epistemological ideas in the macroscopical level. Far from equilibrium they discover an admirable operation of the physical-chemical systems. That is, the discovered the possibility of long range spatiotemporal correlations development when the system lives far from equilibrium. Thus, Prigogine[19] and Nicolis[20] opened a new road towards to the understanding of random fields and statistics, which lead to a non-Gaussian reality. This behavior of nature is called Self-Organization. Prigogine's and Nicoli's self-organized concepts inspired one of the writers of this paper to introduce the self organization theory as basic tool to interpret the dynamics of the space plasmas dynamics [21] as well as seismogenesis [22] as a result of the self organization of Earth's manage-crust system. However Lorenz [23] had discovered the Lorenz's attractor as the weather's self organization process while other scientists had observed the self organization of fluids (e.g. dripping faucet model) or else, verifying the Feigenbaum [24] mathematical scenarios to complexity includes in nonlinear maps or Ordinary Differential Equations – Partial Differential Equations [25]. However, scientists still now prefers to follow the classical theory, namely that macro-cosmos is just the result of fundamental laws which can be traced only at the microscopical level. Therefore, while the supporter of classical reductionistic theory considers the chaos and the self organization macroscopic characteristics that they ought to be the result of the fundamental Lagrangian or the fundamental Hamiltonian of nature, there is an ongoing a different perception. Namely, that macroscopic chaos and complexity not only cannot be explained by the hypothetical microscopic simplicity but they are present also in the microscopic reality.

Therefore, scientists like Nelson [26], Hooft [27], Parisi [28], Beck [29] and others used the complexity concept for the explanation of the microscopic "simplicity", introducing theories like stochastic quantum field theory or chaotic field theory. This new perception started to appear already through the Wilsonian theories of renormalization which showed the multiscale cooperation of the physical reality [30]. At the same time, the multiscale cooperativity goes with the self similarity characters of nature that allows the renormalization process. This leads to the utilization of fractal geometry into the unification of physical theories, as the fractal geometries are characterized by the scaling property which includes the multiscale and self similar character. Scientists like Ord [2], El Naschie [3], Nottale [4] and others, will introduce the idea of fractal geometry into the geometry of space-time, negating the notion of differentiability of physical variables. The fractal geometry is connected to non-commutative geometry since at fractal objects the principle of self similarity negates the notion of the simple geometrical point just like the idea of differentiability. Therefore, the fractal geometry of space-time is leading to the fractal extension of dynamics exploiting the fractal calculus (fractal

integrals- fractal derivatives) [9, 11, 12, 31]. Also, the fractal structure of space-time has intrinsically a stochastic character since a presupposition for determinism is differentiability [4, 11]. Thus, in this way, statistics are unified with dynamics automatically, while the notion of probability obtains a physical substance, characterized as dynamical probabilism. The ontological character of probabilism can be the base for the scientific interpretation of self-organization and ordering principles just as Prigogine [19] had imagined, following Heisenberg's concept. From this point of view, we could say that contemporary physical theory returns to the Aristotetiles point of view as Aristotelianism includes the Newtonian and Democritian mechanistical determinism as one component of the organism like behavior of Nature [34].

Modern evolution of physical theory as it was described previously is highlighted in Tsallis $q$-statistics generalization of the Boltzmann-Gibbs statistics which includes the classical (Gaussian) statistics, as the $q=1$ limit of thermodynamical equilibrium. Far from equilibrium, the statistics of the dynamics follows the $q$-Gaussian generalization of the B-G statistics or other more generalized statistics. At the same time, Tsallis $q$-extension of statistics can be produced by the fractal generalization of dynamics. The traditional scientific point of view is the priority of dynamics over statistics. That is dynamics creates statistics. However for complex system their holistic behaviour does not permit easily such a simplification and division of dynamics and statistics. Tsallis $q$ – statistics and fractal or strange kinetics are two faces of the same complex and holistic (non-reductionist) reality.

Moreover, the Tsallis statistical theory including the Tsallis extension of entropy to what is known as $q$-entropy [1], the fractal generalization of dynamics [6, 7] and the scale extension of relativity theory C [4, 5] are the cornerstones of modern physical theory related with nonlinearity and non-integrability as well as with the non-equilibrium ordering and self organization.

In the following paper we analyze experimental time series extracted from various non-equilibrium dynamical systems and we testify the predictions of Tsallis q-statistics theory by the experimental estimation of Tsallis q-triplet. In section (2) we present useful theoretical concepts presupposed for understanding of the experimental data analysis. In section (3) we describe the methodology for the data analysis. Finally in section (4) we summarize and discuss the results of data analysis.

## 2. Theoretical Concepts

### 2.1 Complexity Theory and the Cosmic Ordering Principle

The conceptual novelty of complexity theory embraces all of the physical reality from equilibrium to non-equilibrium states. This is noticed by Castro as follows: *"…it is reasonable to suggest that there must be a deeper organizing principle, from small to large scales, operating in nature which might be based in the theories of complexity, non-linear dynamics and information theory which dimensions, energy and information are intricately connected." [5].* Tsallis non-extensive statistical theory [1] can be used for a comprehensive description of space plasma dynamics, as recently we became aware of the drastic change of fundamental physical theory concerning physical systems far from equilibrium.

The dynamics of complex systems is one of the most interesting and persisting modern physical problems including the hierarchy of complex and self-organized phenomena such as: intermittent turbulence, fractal structures, long range correlations, far from equilibrium phase transitions, anomalous diffusion – dissipation and strange kinetics, reduction of dimensionality etc [13, 33-39].

More than other scientists, Prigogine, as he was deeply inspired by the arrow of time and the chemical complexity, supported the marginal point of view that the dynamical determinism of physical reality is produced by an underlying ordering process of entirely holistic and probabilistic character at every physical level. If we accept this extreme scientific concept, then we must accept also for complex systems the new point of view, that the classical kinetic is inefficient to describe sufficiently the emerging complex character as the system lives far from equilibrium. However resent evolution of the physical theory centered on non-linearity and fractality shows that the Prigogine point of view was so that much extreme as it was considered at the beginning.

After all, Tsallis $q$–extension of statistics [1] and the fractal extension for dynamics of complex systems as it has been developed by Notalle [4], El Naschie [3], Castro [5], Tarasov [9], Zaslavsky [6], Milovanov [35], El Nabulsi [10], Cresson [11], Coldfain [12], Chen [18], and others scientists, they are the double face of a unified novel theoretical framework, and they constitute the appropriate base for the modern study of non-equilibrium dynamics as the q-statistics is related at its foundation to the underlying fractal dynamics of the non-equilibrium states.

For complex systems near equilibrium the underlying dynamics and the statistics are Gaussian as it is caused by a normal Langevin type stochastic process with a white noise Gaussian component. The normal Langevin stochastic equation corresponds to the probabilistic description of dynamics by the well-known normal Fokker – Planck equation. For Gaussian processes only the moments-cummulants of first and second order are non-zero, while the central limit theorem inhibits the development of long range correlations and macroscopic self-organization, as any kind of fluctuation quenches out exponentially to the normal distribution. Also at equilibrium, the dynamical attractive phase space is practically infinite dimensional as the system state evolves in all dimensions according to the famous ergodic theorem of Boltzmann – Gibbs statistics. However, in Tsallis $q$–statistics even for the case of $q = 1$ (corresponding to Gaussian process) the non-extensive character permits the development of long range correlations produced by equilibrium phase transition multi-scale processes according to the Wilson [30]. From this point of view, the classical mechanics (particles and fields), including also general relativity theory, as well as the quantum mechanics – quantum field theories, all of them are nothing else than a near thermodynamical equilibrium approximation of a wider theory of physical reality, characterized as complexity theory. This theory can be related with a globally acting ordering process which produces the $q$–statistics and the fractal extension of dynamics classical or quantum.

Generally, the experimental observation of a complex system presupposes non-equilibrium process of the physical system which is subjected to observation, even if the system lives thermodynamically near to equilibrium states. Also experimental observation includes discovery and ascertainment of correlations in space and time, as the spatio-temporal correlations are related or they are caused by from the statistical mean values fluctuations. The theoretical interpretation prediction of observations as spatial and temporal correlations – fluctuations is based on statistical theory which relates the microscopic underling dynamics with the macroscopic observations indentified to statistical moments and cumulants. Moreover, it is known that statistical moments and cumulants are related to the underlying dynamics by the derivatives of the partition function ($Z$) to the external source variables ($J$) [40].

From this point of view, the main problem of complexity theory is how to extend the knowledge from thermodynamical equilibrium states to the far from equilibrium physical states. The non-extensive $q$–statistics introduced by Tsallis [1] as the extension of Boltzmann – Gibbs equilibrium statistical theory is the appropriate base for the non-equilibrium extension of complexity theory. The far from equilibrium $q$–statistics can produce the $q$-partition function ($Z_q$) and the corresponding $q$ – moments and cumulants, in correspondence with Boltzmann – Gibbs statistical interpretation of thermodynamics.

The miraculous consistency of physical processes at all levels of physical reality, from the macroscopic to the microscopic level, as well as the inefficiency of existing theories to produce or to predict the harmony and hierarchy of structures inside structures from the macroscopic or the microscopic level of cosmos. This completely supports or justifies such new concepts as that indicated by Castro [5]: *"of a global ordering principle or that indicated by Prigogine, about the becoming before being at every level of physical reality."* The problem however with such beautiful concepts is how to transform them into an experimentally testified scientific theory.

The Feynman path integral formulation of quantum theory after the introduction of imaginary time transformation by the Wick rotation indicates the inner relation of quantum dynamics and statistical mechanics [41, 42]. In this direction it was developed the stochastic and chaotic quantization theory [27-29, 43], which opened the road for the introduction of the macroscopic complexity and self-organization in the region of fundamental quantum field physical theory. The unified character of macroscopic and microscopic complexity is moreover verified by the fact that the $n$ – point Green functions produced by the generating functional $W(J)$ of QFT after the Wick rotation can be transformed to $n$ – point correlation functions produced by the partition function $Z(J)$ of the statistical theory. This indicates in reality the self-organization process underlying the creation and interaction of elementary particles, similarly to the development of correlations in complex systems and classical random fields Parisi [28]. For this reason lattice theory describes simultaneously microscopic and macroscopic complexity [30, 41].

In this way, instead of explaining the macroscopic complexity by a fundamental physical theory such as QFT, Superstring theory, M-theory or any other kind of fundamental theory we become witnesses of the opposite fact, according to what Prigogine was imagining. That is, macroscopic self-organization process and macroscopic complexity install their kingdom in the heart of reductionism and fundamentalism of physical theory. The Renormalizable field theories with the strong vehicle of Feynman diagrams that were used for the description of high energy interactions or the statistical theory of critical phenomena and the nonlinear dynamics of plasmas [44] lose their efficiency when the complexity of the process scales up [30].

Many scientist as Chang [34], Zelenyi [33], Milovanov [35], Ruzmaikin [36], Abramenko [39], Lui [45], Pavlos [13], in their studies indicate the statistical non-extensivity as well as the multi-scale, multi-fractal and anomalous – intermittent character of fields and particles in the space plasmas and other complex systems far from equilibrium. These results verify the concept that space plasmas and other complex systems dynamics are part of the more general theory of fractal dynamics which has been developed rapidly the last years. Fractal dynamics are the modern fractal extension of physical theory in every level. On the other side the fractional generalization of modern physical theory is based on fractional calculus: fractional derivatives or integrals or fractional calculus of scalar or vector fields and fractional functional calculus [9, 18]. It is very impressive the efficiency of fractional calculus to describe complex and far from equilibrium systems which display scale-invariant properties, turbulent dissipation and long range correlations with memory preservation, while these characteristics cannot be illustrated by using traditional analytic and differentiable functions, as well as, ordinary differential operators. Fractional calculus permits the fractal generalization of Lagrange – Hamilton theory of Maxwell equations and Magnetohydrodynamics, the Fokker – Planck equation Liouville theory and BBGKI hierarchy, or the fractal generalization of QFT and path integration theory [9-12, 18].

According to the fractal generalization of dynamics and statistics we conserve the continuity of functions but abolish their differentiable character based on the fractal calculus which are the non-differentiable generalization of differentiable calculus. At the same time the deeper physical meaning of fractal calculus is the unification of microscopic and macroscopic dynamical theory at the base of the space – time fractality [3, 4, 18, 46-48]. Also, the space-time is related to the fractality – multi-fractality of the dynamical phase – space, whish can be manifested as non-equilibrium complexity and self-organization.

Moreover fractal dynamics leads to a global generalization of physical theory as it can be related with the infinite dimension Cantor space, as the microscopic essence of physical space – time, the non-commutative geometry and non-commutative Clifford manifolds and Clifford algebra, or the p-adic physics [3, 5, 10, 49, 50]. According to these new concepts introduced the last two decades at every level of physical reality we can describe in physics complex structure which cannot be reduced to underlying simple fundamental entities or underlying simple fundamental laws. Also, the non-commutative character of physical theory and geometry indicates [50, 51] that the scientific observation is nothing more than the observation of undivided complex structures in every level. Cantor was the founder of the fractal physics creating fractal sets by contraction of the homogenous real number set, while on the other side the set of real numbers can be understood as the result of the observational coarse graining [49, 52]. From a philosophical point of view the mathematical forms are nothing else than self-organized complex structures of the mind-brain, in self-consistency with all the physical reality. On the other side, the generalization of Relativity theory to scale relativity by Nottale [4] or Castro [5] indicates the unification of microscopic and macroscopic dynamics through the fractal generalization of dynamics.

After all, we conjecture that the macroscopic self-organization related with the novel theory of complex dynamics, as they can be observed at far from equilibrium dynamical physical states, are the macroscopic emergence result of the microscopic complexity which can be enlarged as the system arrives at bifurcation or far from equilibrium critical points. That is, far from equilibrium the observed physical self-organization manifests the globally active ordering principle to be in priority from local interactions processes. We could conjecture that is not far from thruth the concept that local interactions themselves are nothing else than local manifestation of the holistically active ordering principle. That is what until now is known as fundamental lows is the equilibrium manifestation or approximation of the new and globally active ordering principle. This concept can be related with the fractal generalization of dynamics which is indentified with the dynamics of correlations supported by Prigogine [19], Nicolis [20] and Balescu [53], as the generalization of Newtonian theory. This conjecture concerning the fractal unification of macroscopic and microscopic dynamics at can be strongly supported by the Tsallis nonextensive q-statistics theory which is verified almost everywhere from the microscopic to the macroscopic level [1, 5]. From this point of view it is reasonable to support that the q-statistics and the fractal generalization of space plasma dynamics is the appropriate framework for the description of their non-equilibrium complexity.

## 2.2 Chaotic Dynamics and Statistics

The macroscopic description of complex systems can be approximated by non-linear partial differential equations of the general type:

$$\frac{\partial \vec{U}(\vec{x},t)}{\partial t} = \vec{F}(\vec{u}, \vec{\lambda}) \qquad (1)$$

where $u$ belongs to a infinite dimensional state (phase) space which is a Hilbert functional space. Among the various control parameters, the plasma Reynold number is the one which controls the quiet static or the turbulent plasma states. Generally the control parameters measure the distance from the thermodynamical equilibrium as well as the critical or bifurcation points of the system for given and fixed values, depending upon the global mathematical structure of the dynamics. As the system passes its bifurcation points a rich variety of spatio-temporal patterns with distinct topological and dynamical profiles can be emerged such as: limit cycles or torus, chaotic or strange attractors, turbulence, Vortices, percolation states and other kinds of complex spatiotemporal structures [34, 48, 54-62].

Generally chaotic solutions of the mathematical system (1) transform the deterministic form of equation (1) to a stochastic non-linear stochastic system:

$$\frac{\partial \vec{u}}{\partial t} = \vec{\Phi}(\vec{u}, \vec{\lambda}) + \vec{\delta}(\vec{x}, t) \qquad (2)$$

where $\vec{\delta}(\vec{x}, t)$ corresponds to the random force fields produced by strong chaoticity [42].

The non-linear mathematical systems (1-2) include mathematical solutions which can represent plethora of non-equilibrium physical states included in mechanical, electromagnetic or chemical and other physical systems which are study here.

The random components ($\delta(\vec{x}, t)$) are related to the BBGKY hierarchy:

$$\frac{\partial f_q}{\partial t} = [H_q, f_a] + S_q, q = 1, 2, ..., N \qquad (3)$$

where $f_q$ is the $q$-particle distribution function, $H_q$ is the $q$-th approximation of the Hamiltonian $q$-th correlations and $S_q$ is the statistical term including correlations of higher than q-orders [44, 17].

The non-linear mathematical systems (1,2) correspond to the new science known today as complexity science. This new science has a universal character, including an unsolved scientific and conceptual controversy which is continuously spreading in all directions of the physical reality and concerns the integrability or computability of the dynamics [63]. This universality is something supported by many scientists after the Poincare discovery of chaos and its non-integrability as is it shown in physical sciences by the work of Prigogine, Nicolis, Yankov and others [19, 21, 63] in reality. Non-linearity and chaos is the top of a hidden mountain including new physical and mathematical concepts such as fractal calculus, p-adic physical theory, non-commutative geometry, fuzzy anomalous topologies fractal space-time etc [3, 5, 9-12, 6, 18, 49-51]. These new mathematical concepts obtain their physical power when the physical system lives far from equilibrium.

After this and, by following the traditional point of view of physical science we arrive at the central conceptual problem of complexity science. That is, how is it possible that the local interactions in a spatially distributed physical system can cause long range correlations or how they can create complex spatiotemporal coherent patterns as the previous non-linear mathematical systems reveal, when they are solved arithmetically, or in situ observations reveal in space plasma systems. For non-equilibrium physical systems the above questions make us to ask how the development of complex structures and long range spatio-temporal correlations can be explained and described by local interactions of particles and fields. At a first glance the problem looks simple supposing that it can be explained by the self-consistent particle-fields classical interactions. However the existed rich phenomenology of complex non-equilibrium phenomena reveals the non-classical and strange character of the universal non-equilibrium critical dynamics [34, 40].

In the following and for the better understanding of the new concepts we follow the road of non-equilibrium statistical theory [34, 64]

The stochastic Langevin equations (65, 10, 82) can take the general form:

$$\frac{\partial u_i}{\partial t} = -\Gamma(\vec{x})\frac{\delta H}{\delta u_i(\vec{x},t)} + \Gamma(\vec{x})n_i(\vec{x},t) \qquad (4)$$

where $H$ is the Hamiltonian of the system, $\delta H / \delta u_i$ its functional derivative, $\Gamma$ is a transport coefficient and $n_i$ are the components of a Gaussian white noise:

$$\left. \begin{array}{l} <n_i(\vec{x},t)> = 0 \\ <n_i(\vec{x},t)n_j(\vec{x}',t')> = 2\Gamma(\vec{x})\delta_{ij}\delta(\vec{x}-\vec{x}')\delta(t-t') \end{array} \right\} \qquad (5)$$

[34, 64, 65]. The above stochastic Langevin Hamiltonian equation (4) can be related to a probabilistic Fokker – Planck equation [34]:

$$\frac{1}{\Gamma(\vec{x})}\frac{\partial P}{\partial t} = \frac{\delta}{\delta \vec{u}} \cdot \left( \frac{\delta H}{\delta \vec{u}}P + \frac{\delta}{\delta \vec{u}}[\Gamma(\vec{x})P] \right) \qquad (6)$$

where $P = P(\{u_i(\vec{x},t)\},t)$ is the probability distribution function of the dynamical configuration $\{u_i(\vec{x},t)\}$ of the system at time $t$. The solution of the Fokker – Planck equation can be obtained as a functional path integral in the state space $\{u_i(\vec{x})\}$:

$$P(\{u_i(\vec{x})\},t) \simeq \int \Delta\vec{Q}\exp(-S)P_0(\{u_i(\vec{x})\},t_0) \qquad (7)$$

where $P_0(\{u_i(\vec{x})\},t_0)$ is the initial probability distribution function in the extended configuration state space and $S = i\int Ldt$ is the stochastic action of the system obtained by the time integration of it's stochastic Lagrangian (L) [5, 34]. The stationary solution of the Fokker – Planck equation corresponds to the statistical minimum of the action and corresponds to a Gaussian state:

$$P(\{u_i\}) \sim \exp\left[-(1/\Gamma)H(\{u_i\})\right] \qquad (8)$$

The path integration in the configuration field state space corresponds to the integration of the path probability for all the possible paths which start at the configuration state $\vec{u}(\vec{x},t_0)$ of the system and arrive at the final configuration state $\vec{u}(\vec{x},t)$. Langevin and F-P equations of classical statistics include a hidden relation with Feynman path integral formulation of QM [28, 34, 42]. The F-P equation can be transformed to a Schrödinger equation:

$$i\frac{d}{dt}\hat{U}(t,t_0) = \hat{H}\cdot\hat{U}(t,t_0) \qquad (9)$$

by an appropriate operator Hamiltonian extension $H(u(\vec{x},t)) \Rightarrow \hat{H}(\hat{u}(\vec{x},t))$ of the classical function $(H)$ where now the field $(u)$ is an operator distribution [34, 65]. From this point of view, the classical stochasticity of the macroscopic Langevin process can be considered as caused by a macroscopic quandicity revealed by the complex system as the F-K probability distribution $P$ satisfies the quantum relation:

$$P(u,t|u,t_0) = \langle u|\hat{U}(t,t_0)|u_0\rangle \qquad (10)$$

This generalization of classical stochastic process as a quantum process could explain the spontaneous development of long-range correlations at the macroscopic level as an enlargement of the quantum entanglement character at critical states of complex systems. This interpretation is in faithful agreement with the introduction of complexity in sub-quantum processes and the chaotic – stochastic quantization of field theory [27-29, 43], as well as with scale relativity principles [4, 5, 48] and fractal extension of dynamics [3, 9, 10-12, 18] or the older Prigogine self-organization theory [19]. Here, we can argue in addition to previous description that quantum mechanics is subject gradually to a fractal generalization [5, 9, 10-12]. The fractal generalization of QM-QFT drifts along also the tools of quantum theory into the correspondent generalization of RG theory or path integration and Feynman diagrams. This generalization implies also the generalization of statistical theory as the new road for the unification of macroscopic and microscopic complexity.

If $P[\vec{u}(\vec{x},t)]$ is the probability of the entire field path in the field state space of the distributed system, then we can extend the theory of generating function of moments and cumulants for the probabilistic description of the paths [15, 59]. The n-point field correlation functions (n-points moments) can be estimated by using the field path probability distribution and field path (functional) integration:

$$\langle u(\vec{x}_1,t_1)u(\vec{x}_2,t_2)...u(x_n,t_n)\rangle = \int \Delta \vec{u} P[\vec{u}(\vec{x},t)] u(\vec{x}_1,t_1)...u(\vec{x}_n,t_n) \qquad (11)$$

For Gaussian random processes which happen to be near equilibrium the $n$ – th point moments with $n > 2$ are zero, correspond to Markov processes while far from equilibrium it is possible non-Gaussian (with infinite nonzero moments) processes to be developed. According to Haken [15] the characteristic function (or generating function) of the probabilistic description of paths:

$$[u(x,t)] \equiv \left(u(\vec{x}_1,t_1), u(\vec{x}_2,t_2),...,u(\vec{x}_n,t_n)\right) \qquad (12)$$

is given by the relation:

$$\Phi_{path}\left(j_1(t_1), j_2(t_2),...,j_n(t_n)\right) = \left\langle \exp i \sum_{i=1}^{N} j_i u(\vec{x}_i,t_i) \right\rangle_{path} \qquad (13)$$

while the path cumulants $K_s(t_{a_1}...t_{a_s})$ are given by the relations:

$$\Phi_{path}\left(j_1(t_1), j_2(t_2),...,j_n(t_n)\right) = \exp\left\{\sum_{s=1}^{\infty} \frac{i^s}{s!} \sum_{a_1,...,a_s=1}^{n} K_s(t_{a_1}...t_{a_s}) \cdot j_{a_1}...j_{a_s}\right\} \qquad (14)$$

and the $n$ – point path moments are given by the functional derivatives:

$$\langle u(\vec{x}_1,t_1), u(\vec{x}_2,t_2),...,u(\vec{x}_n,t_n)\rangle = \left(\delta^n \Phi(\{j_i\})/\delta j_1...\delta j_n\right)t\{j_i\} = 0 \qquad (15)$$

For Gaussian stochastic field processes the cumulants except the first two vanish $(k_3 = k_4 = ...0)$. For non-Gaussian processes it is possible to be developed long range correlations as the cummulants of higher than two order are non-zero [15]. This is the deeper meaning of non-equilibrium self-organization and ordering of complex systems. The characteristic function of the dynamical stochastic field system is related to the partition functions of its statistical description, while the cumulant development and multipoint moments generation can be related with the BBGKY statistical hierarchy of the statistics as well as with the Feynman diagrams approximation of the stochastic field system [40, 66]. For dynamical systems near equilibrium only the second order cumulants is non-vanishing, while far from equilibrium field fluctuations with higher – order non-vanishing cumulants can be developed.

Finally, we can understand how the non-linear dynamics correspond to self-organized states as the high-order (infinite) non-vanishing cumulants can produce the non-integrability of the dynamics. From this point of view the linear or non-linear instabilities of classical kinetic theory are inefficient to produce the non-Gaussian, holistic (non-local) and self-organized complex character of non-equilibrium dynamics. That is, far from equilibrium complex states can be developed including long range correlations of field and particles with non-Gaussian distributions of their dynamic variables. As we show in the next section such states such states reveal the necessity of new theoretical tools for their understanding which are much different from the classical linear or non-linear approximation of kinetic theory.

## 2.3 Strange attractors and Self-Organization

When the dynamics is strongly nonlinear then for the far from equilibrium processes it is possible to be created strong self-organization and intensive reduction of dimensionality of the state space, by an attracting low dimensional set with parallel development of long range correlations in space and time. The attractor can be periodic (limit cycle, limit m-torus), simply chaotic (mono-fractal) or strongly chaotic with multiscale and multifractal profile as well as attractors with weak chaotic profile known as SOC states. This spectrum of distinct dynamical profiles can be obtained as distinct critical points (critical states) of the nonlinear dynamics, after successive bifurcations as the control parameters change. The fixed points can be estimated by using a far from equilibrium renormalization process as it was indicated by Chang [34].

From this point of view phase transition processes can be developed by between different critical states, when the order parameters of the system are changing. The far from equilibrium development of chaotic (weak or strong) critical states include long range correlations and multiscale internal self organization. Now, these far from equilibrium self organized states, the equilibrium BG statistics and BG entropy, are transformed and replaced by the Tsallis extension of $q$-statistics and Tsallis entropy. The extension of renormalization group theory and critical dynamics, under the $q$-extension of partition function, free energy and path integral approach has been also indicated [13, 66, 67]. The multifractal structure of the chaotic attractors can be described by the generalized Rényi fractal dimensions:

$$D_{\bar{q}} = \frac{1}{\bar{q}-1} \lim_{\lambda \to 0} \frac{\log \sum_{i=1}^{N\lambda} p_i^{\bar{q}}}{\log \lambda}, \tag{16}$$

where $p_i \sim \lambda^{\alpha(i)}$ is the local probability at the location ($i$) of the phase space, $\lambda$ is the local size of phase space and $a(i)$ is the local fractal dimension of the dynamics. The Rényi $\bar{q}$ numbers (different from the $q$-index of Tsallis statistics) take values in the entire region $(-\infty, +\infty)$ of real numbers. The spectrum of distinct local fractal dimensions $\alpha(i)$ is given by the estimation of the function $f(\alpha)$ [68, 69] for which the following relations hold:

$$\sum p_i^{\bar{q}} = \int d\alpha' p(\alpha') \lambda^{-f(\alpha')} d\alpha' \tag{17}$$

$$\tau(\bar{q}) \equiv (\bar{q}-1) D_{\bar{q}} \stackrel{\min}{=} \bar{q}\alpha - f(\alpha) \tag{18}$$

$$a(\bar{q}) = \frac{d[\tau(\bar{q})]}{d\bar{q}} \tag{19}$$

$$f(\alpha) = \bar{q}\alpha - \tau(\bar{q}), \tag{20}$$

The physical meaning of these magnitudes included in relations (17-20) can be obtained if we identify the multifractal attractor as a thermodynamical object, where its temperature ($T$), free energy ($F$), entropy ($S$) and internal energy ($U$) are related to the properties of the multifractal attractor as follows:

$$\left. \begin{array}{ll} \bar{q} \Rightarrow \frac{1}{T}, & \tau(\bar{q}) = (\bar{q}-1)D_q \Rightarrow F \\ \alpha \Rightarrow U, & f(\alpha) \Rightarrow S \end{array} \right\} \tag{21}$$

This correspondence presents the relations (19-21) as a thermodynamical Legendre transform [70]. When $\bar{q}$ increases to infinite $(+\infty)$, which means, that we freeze the system ($T_{(q=+\infty)} \to 0$), then the trajectories (fluid lines) are closing on the attractor set, causing large probability values at regions of low fractal dimension, where $\alpha = \alpha_{\min}$ and $D_{\bar{q}} = D_{-\infty}$. Oppositely, when $\bar{q}$ decreases to infinite ($-\infty$), that is we warm up the system ($T_{(q=-\infty)} \to 0$) then the trajectories are spread out at regions of high fractal dimension ($\alpha \Rightarrow \alpha_{\max}$). Also for $\bar{q}' > \bar{q}$ we have $D_{\bar{q}'} < D_{\bar{q}}$ and $D_{\bar{q}} \Rightarrow D_{+\infty}(D_{-\infty})$ for $\alpha \Rightarrow \alpha_{\min}(\alpha_{\max})$ correspondingly. However, the above description presents only a weak or limited analogy between multifractal and thermodynamical objects. The real thermodynamical character of the multifractal objects and multiscale dynamics was discovered after the definition by Tsallis [1] of the $q-$entropy related with the $q-$statistics as it is summarized in the next section (2.6).

## 2.4 Intermittent Turbulence

According to previous description dissipative nonlinear dynamics can produce self-organization and long range correlations in space and time. In this case we can imagine the mirroring relationship between the phase space multifractal attractor and the corresponding multifractal turbulence dissipation process of the dynamical system in the physical space. Multifractality and multiscaling interaction, chaoticity and mixing or diffusion (normal or anomalous), all of them can be manifested in both the state (phase) space and the physical (natural) space as the mirroring of the same complex dynamics. We could say that turbulence is for complexity theory, what the blackbody radiation was for quantum theory, as all previous characteristics can be observed in turbulent states. The theoretical description of turbulence in the physical space is based upon the concept of the invariance of the HD or MHD equations upon scaling transformations to the space-time variables ($\vec{X}, t$) and velocity ($\vec{U}$):

$$\vec{X}' = \lambda \vec{X}, \ \vec{U}' = \lambda^{a/3}\vec{U}, t' = \lambda^{1-a/3}t \qquad (22)$$

and corresponding similar scaling relations for other physical variables [45, 74]. Under these scale transformations the dissipation rate of turbulent kinetic or dynamical field energy $E_n$ (averaged over a scale $l_n = l_o \delta_n = R_o \delta_n$) rescales as $\varepsilon_n$:

$$\varepsilon_n \sim \varepsilon_0 (l_n \backslash l_0)^{\alpha-1} \qquad (23)$$

Kolmogorov assumes no intermittency as the locally averaged dissipation rate [71], in reality a random variable, is independent of the averaging domain. This means in the new terminology of Tsallis theory that Tsallis $q$-indices satisfy the relation $q = 1$ for the turbulent dynamics in the three dimensional space. That is the multifractal (intermittency) character of the HD or the MHD dynamics consists in supposing that the scaling exponent $\alpha$ included in relations (22, 23) takes on different values at different interwoven fractal subsets of the $d-$dimensional physical space in which the dissipation field is embedded. The exponent $\alpha$ and for values $a < d$ is related with the degree of singularity in the field's gradient ($\frac{\partial A(x)}{\partial x}$) in the $d-$dimensional natural space [72]. The gradient singularities cause the anomalous diffusion in physical or in phase space of the dynamics. The total dissipation occurring in a $d-$dimensional space of size $l_n$ scales also with a global dimension $D_{\bar{q}}$ for powers of different order $\bar{q}$ as follows:

$$\sum_n \varepsilon_n^{\bar{q}} l_n^d \sim l_n^{(\bar{q}-1)D_{\bar{q}}} = l_n^{\tau(\bar{q})} \qquad (24)$$

Supposing that the local fractal dimension of the set $dn(a)$ which corresponds to the density of the scaling exponents in the region ($\alpha, \alpha + d\alpha$) is a function $f_d(a)$ according to the relation:

$$dn(\alpha) \sim \ln^{-f_d(\alpha)} da \qquad (25)$$

where $d$ indicates the dimension of the embedding space, then we can conclude the Legendre transformation between the mass exponent $\tau(\bar{q})$ and the multifractal spectrum $f_d(a)$:

$$\left.\begin{array}{l} f_d(a) = a\bar{q} - (\bar{q}-1)(D_{\bar{q}} - d + 1) + d - 1 \\ a = \dfrac{d}{d\bar{q}}[(\bar{q}-1)(D_{\bar{q}} - d + 1)] \end{array}\right\} \quad (26)$$

For linear intersections of the dissipation field, that is $d = 1$ the Legendre transformation is given as follows:

$$f(a) = a\bar{q} - \tau(\bar{q}), \quad a = \frac{d}{d\bar{q}}[(q-1)D_q] = \frac{d}{d\bar{q}}\tau(\bar{q}), \quad \bar{q} = \frac{df(a)}{da} \quad (27)$$

The relations (25-27) describe the multifractal and multiscale turbulent process in the physical state. The relations (16-21) describe the multifractal and multiscale process on the attracting set of the phase space. From this physical point of view, we suppose the physical identification of the magnitudes $D_{\bar{q}}, a, f(a)$ and $\tau(\bar{q})$ estimates in the physical and the corresponding phase space of the dynamics. By using experimental timeseries we can construct the function $D_{\bar{q}}$ of the generalized Rényi $d-$dimensional space dimensions, while the relations (26) allow the calculation of the fractal exponent ($a$) and the corresponding multifractal spectrum $f_d(a)$. For homogeneous fractals of the turbulent dynamics the generalized dimension spectrum $D_{\bar{q}}$ is constant and equal to the fractal dimension, of the support [71]. Kolmogorov [73] supposed that $D_{\bar{q}}$ does not depend on $\bar{q}$ as the dimension of the fractal support is $D_q = 3$. In this case the multifractal spectrum consists of the single point ($a = 1$ and $f(1) = 3$). The singularities of degree ($a$) of the dissipated fields, fill the physical space of dimension $d$ with a fractal dimension $F(a)$, while the probability $P(a)da$, to find a point of singularity ($a$) is specified by the probability density $P(a)da \sim \ln^{d-F(a)}$. The filling space fractal dimension $F(a)$ is related with the multifractal spectrum function $f_d(a) = F(a) - (d-1)$, while according to the distribution function $\Pi_{dis}(\varepsilon_n)$ of the energy transfer rate associated with the singularity $a$ it corresponds to the singularity probability as $\Pi_{dis}(\varepsilon_n)d\varepsilon_n = P(a)da$ [72].

Moreover the partition function $\sum_i P_i^{\bar{q}}$ of the Rényi fractal dimensions estimated by the experimental timeseries includes information for the local and global dissipation process of the turbulent dynamics as well as for the local and global dynamics of the attractor set, as it is transformed to the partition function $\sum_i P_i^q = Z_q$ of the Tsallis q-statistic theory.

## 2.5 Fractal generalization of dynamics.

Fractal integrals and fractal derivatives are related with the fractal contraction transformation of phase space as well as contraction transformation of space time in analogy with the fractal contraction transformation of the Cantor set [31, 52]. Also, the fractal extension of dynamics includes an extension of non-Gaussian scale invariance, related to the multiscale coupling and non-equilibrium extension of the renormalization group theory [6]. Moreover Tarasov [9], Coldfain [12], Cresson [11], El-Nabulsi [10] and other scientists generalized the classical or quantum dynamics in a continuation of the original break through of El-Naschie [3], Nottale [4], Castro [5] and others concerning the fractal generalization of physical theory.

According to Tarasov [9] the fundamental theorem of Riemann – Liouville fractional calculus is the generalization of the known integer integral – derivative theorem as follows:

if
$$F(x) = {}_aI_x^a f(x) \qquad (28)$$

then
$$_aD_x^a F(x) = f(x) \qquad (29)$$

where ${}_aI_x^a$ is the fractional Riemann – Liouville according to:

$$_aI_x^a f(x) \equiv \frac{1}{\Gamma(a)} \int_a^x \frac{f(x')dx'}{(x-x')^{1-a}} \qquad (30)$$

and ${}_aD_x^a$ is the Caputo fractional derivative according to:

$$_aD_x^a F(x) = {}_aI_x^{n-a} D_x^n F(x) =$$
$$= \frac{1}{\Gamma(n-a)} \int_a^x \frac{dx'}{(x-x')^{1+a-n}} \frac{d^nF(x)}{dx^n} \qquad (31)$$

for $f(x)$ a real valued function defined on a closed interval $[a,b]$.

In the next we summarize the basic concepts of the fractal generalization of dynamics as well as the fractal generalization of Liouville and MHD theory following Tarasov [9]. According to previous descriptions, the far from equilibrium dynamics includes fractal or multi-fractal distribution of fields and particles, as well as spatial fractal temporal distributions. This state can be described by the fractal generalization of classical theory: Lagrange and Hamilton equations of dynamics, Liouville theory, Fokker Planck equations and Bogoliubov hierarchy equations. In general, the fractal distribution of a physical quantity ($M$) obeys a power law relation:

$$M_D \sim M_0 \left(\frac{R}{R_0}\right)^D \qquad (32)$$

where ($M_D$) is the fractal mass of the physical quantity ($M$) in a ball of radius ($R$) and ($D$) is the distribution fractal dimension. For a fractal distribution with local density $\rho(\vec{x})$ the fractal generalization of Euclidean space integration reads as follows:

$$M_D(W) = \int_W \rho(x) dV_D \qquad (33)$$

where
$$dV_D = C_3(D,\vec{x}) dV_3 \qquad (34)$$

and
$$C_3(D,\vec{x}) = \frac{2^{3-D}\Gamma(3/2)}{\Gamma(D/2)} |\vec{x}|^{D-3} \qquad (35)$$

Similarly the fractal generalization of surface and line Euclidean integration is obtained by using the relations:

$$dS_d = C_2(d,\vec{x}) dS_2 \qquad (36)$$

$$C_2(d,\vec{x}) = \frac{2^{2-d}}{\Gamma(d/2)} |\vec{x}|^{d-2} \qquad (37)$$

for the surface fractal integration and

$$dl_\gamma = C_1(\gamma,\vec{x}) dl_1 \qquad (38)$$

$$C_1(\gamma, \vec{x}) = \frac{2^{1-\gamma}\Gamma(1/2)}{\Gamma(\gamma/2)}|\vec{x}|^{\gamma-1} \tag{39}$$

for the line fractal integration. By using the fractal generalization of integration and the corresponding generalized Gauss's and Stoke's theorems we can transform fractal integral laws to fractal and non-local differential laws. The fractional generalization of classical dynamics (Hamilton Lagrange and Liouville theory) can be obtained by the fractional generalization of phase space quantative description. For this we use the fractional power of coordinates:

$$X^a = \text{sgn}(x)|x|^a \tag{40}$$

where $\text{sgn}(x)$ is equal to +1 for $x \geq 0$ and equal to -1 for $x < 0$.

The fractional measure $M_a(B)$ of a $n$-dimension phase space region $(B)$ is given by the equation:

$$M_a(B) = \int_B g(a) d\mu_a(q,p) \tag{41}$$

where $d\mu_a(q,p)$ is a phase space volume element:

$$d\mu_a = \Pi \frac{dq_K^a \wedge dp_K^a}{[a\Gamma(a)]^2} \tag{42}$$

where $g(a)$ is a numerical multiplier and $dq_K^a \wedge dp_K^a$ means the wedge product.

The fractional Hamilton's approach can be obtained by the fractal generalization of the Hamilton action principle:

$$S = \int [pq - H(t,p,q)]dt \tag{43}$$

The fractal Hamilton equations:

$$\left(\frac{dq}{at}\right)^q = \Gamma(2-a) p^{a-1} D_p^a H \tag{44}$$

$$D_t^a p = -D_q^a H \tag{45}$$

while the fractal generalization of the Lagrange's action principle:

$$S = \int L(t,q,u)dt \tag{46}$$

Corresponds to the fractal Lagrange equations:

$$D_q^a L - \Gamma(2-a) D_t^a \left[D_U^a L\right]_{U=\dot{q}} = 0 \tag{47}$$

Similar fractal generalization can be obtained for dissipative or non-Hamiltonian systems [9]. The fractal generalization of Liouville equation is given also as:

$$\frac{\partial \tilde{p}_N}{\partial t} = L_N \tilde{p}_N \tag{48}$$

where $\tilde{p}_N$ and $L_N$ are the fractal generalization of probability distribution function and the Liouville operator correspondingly. The fractal generalization of Bogoliubov hierarchy can be obtained by using the fractal Liouville equation as well as the fractal Fokker Planck hydrodynamical - magnetohydrodynamical approximations.

The fractal generalization of classical dynamical theory for dissipative systems includes the non-Gaussian statistics as the fractal generalization of Boltzmann – Gibbs statistics.

Finally the far from equilibrium statistical mechanics can be obtained by using the fractal extension of the path integral method. The fractional Green function of the dynamics is given by the fractal generalization of the path integral:

$$K_a\left(x_f, t_f; x_i, t_i\right) \simeq \int_{x_i}^{x_f} D[x_a(\tau)] \exp\left[\frac{i}{h} S_a(\gamma)\right]$$
$$\simeq \sum_{\{\gamma\}} \exp\left[\frac{i}{h} S_a(\gamma)\right]$$
(49)

where $K_a$ is the probability amplitude (fractal quantum mechanics) or the two point correlation function (statistical mechanics), $D[x_a(\tau)]$ means path integration on the sum $\{\gamma\}$ of fractal paths and $S_a(\gamma)$ is the fractal generalization of the action integral:

$$S_a[\gamma] = \frac{1}{\Gamma(a)} \int_{x_i}^{x_f} L\left(D_\gamma^a q(\tau), \tau\right)(t-\tau)^{a-1} d\tau$$
(50)

## 2.6 The Highlights of Tsallis Theory

As we show in the next sections of this study, everywhere in space plasmas we can ascertain the presence of Tsallis statistics. This discovery is the continuation of a more general ascertainment of Tsallis q-extensive statistics from the macroscopic to the microscopic level [1].

In our understanding the Tsallis theory, more than a generalization of thermodynamics for chaotic and complex systems, or a non-equilibrium generalization of B-G statistics, can be considered as a strong theoretical vehicle for the unification of macroscopic and microscopic physical complexity. From this point of view Tsallis statistical theory is the other side of the modern fractal generalization of dynamics while its essence is nothing else than the efficiency of self-organization and development of long range correlations of coherent structures in complex systems.

From a general philosophical aspect, the Tsallis q-extension of statistics can be identified with the activity of an ordering principle in physical reality, which cannot be exhausted with the local interactions in the physical systems, as we noticed in previous sections.

### 2.6.1 The non-extensive entropy ($S_q$).

It was for first time that Tsallis [1], inspired by multifractal analysis, conceived that the Boltzmann – Gibbs entropy:

$$S_{BG} = -K \sum p_i \ln p_i, \quad i = 1, 2, ..., N$$
(51)

is inefficient to describe all the complexity of non-linear dynamical systems. The Boltzmann – Gibbs statistical theory presupposes ergodicity of the underlying dynamics in the system phase space. The complexity of dynamics which is far beyond the simple ergodic complexity, it can be described by Tsallis non-extensive statistics, based on the extended concept of $q$ – entropy:

$$S_q = k\left(1 - \sum_{i=1}^{N} p_i^q\right) / (q-1)$$
(52)

for discrete state space or

$$S_q = k\left[1 - \int [p(x)]^q dx\right] / (q-1)$$
(53)

for continuous state space.
For a system of particles and fields with short range correlations inside their immediate neighborhood, the Tsallis $q$ – entropy $S_q$ asymptotically leads to Boltzmann – Gibbs entropy ($S_{BG}$) corresponding to the value of $q=1$. For probabilistically dependent or correlated systems $A, B$ it can be proven that:

$$S_q(A+B) = S_q(A) + S_q(B/A) + (1-q)S_q(A)S_q(B/A)$$
$$= S_q(B) + S_q(A/B) + (1-q)S_q(B)S_q(A/B) \quad (54)$$

where $S_q(A) \equiv S_q(\{p_i^A\}), S_q(B) \equiv Sq(\{p_i^B\}), S_q(B/A)$ and $S_q(A/B)$ are the conditional entropies of systems $A, B$ [15]. When the systems are probabilistically independent, then relation (54) is transformed to:

$$S_q(A+B) = S_q(A) + S_q(B) + (1-q)S_q(A)S_q(B) \quad (55)$$

The dependent (independent) property corresponds to the relation:

$$p_{ij}^{A+B} \neq p_i^A p_j^B \left( p_{ij}^{A+B} = p_i^A p_j^B \right) \quad (56)$$

Comparing the Boltzmann – Gibbs ($S_{BG}$) and Tsallis ($S_q$) entropies, we conclude that for non-existence of correlations $S_{BG}$ is extensive whereas $S_q$ for $q \neq 1$ is non-extensive. In contrast, for global correlations, large regions of phase – space remain unoccupied. In this case $S_q$ is non-extensive either $q=1$ or $q \neq 1$.

### 2.6.2 The $q$ – extension of statistics and Thermodynamics.

Non-linearity can manifest its rich complex dynamics as the system is removed far from equilibrium. The Tsallis $q$ – extension of statistics is indicated by the non-linear differential equation $dy/dx = y^q$. The solution of this equation includes the $q$ – extension of exponential and logarithmic functions:

$$e_q^x = [1+(1-q)x]^{1/(1-q)} \quad (57)$$
$$\ln_q x = (x^{1-q} - 1)/(1-q) \quad (58)$$

and

$$p_{opt}(x) = e_q^{-\beta_q[f(x)-F_q]} / \int dx' e_q^{-\beta_q[f(x')-F_q]} \quad (59)$$

for more general $q$ – constraints of the forms $\langle f(x) \rangle_q = F_q$. In this way, Tsallis $q$ – extension of statistical physics opened the road for the $q$ – extension of thermodynamics and general critical dynamical theory as a non-linear system lives far from thermodynamical equilibrium. For the generalization of Boltzmann-Gibbs nonequilibrium statistics to Tsallis nonequilibrium q-statistics we follow Binney [40]. In the next we present q-extended relations, which can describe the non-equilibrium fluctuations and $n$ – point correlation function ($G$) can be obtained by using the Tsallis partition function $Z_q$ of the system as follows:

$$G_q^n(i_1, i_2, ..., i_n) \equiv \langle s_{i_1}, s_{i_2}, ..., s_{i_n} \rangle_q = \frac{1}{z} \frac{\partial^n Z_q}{\partial j_{i_1} \cdot \partial j_{i_2} ... \partial j_{i_n}} \quad (60)$$

Where $\{s_i\}$ are the dynamical variables and $\{j_i\}$ their sources included in the effective – Lagrangian of the system. Correlation (Green) equations (60) describe discrete variables, the $n$ – point correlations for continuous distribution of variables (random fields) are given by the functional derivatives of the functional partition as follows:

$$G_q^n(\vec{x}_1, \vec{x}_2, ..., \vec{x}_n) \equiv \langle \varphi(\vec{x}_1)\varphi(\vec{x}_2)...\varphi(\vec{x}_n) \rangle_q = \frac{1}{Z} \frac{\delta}{\delta J(\vec{x}_1)} ... \frac{\delta}{\delta J(\vec{x}_n)} Z_q(J) \quad (61)$$

where $\varphi(\vec{x})$ are random fields of the system variables and $j(\vec{x})$ their field sources. The connected $n$ – point correlation functions $G_i^n$ are given by:

$$G_q^n(\vec{x}_1, \vec{x}_2, ..., \vec{x}_n) \equiv \frac{\delta}{\delta J(\vec{x}_1)} ... \frac{\delta}{\delta J(\vec{x}_n)} \log Z_q(J) \qquad (62)$$

The connected $n$ – point correlations correspond to correlations that are due to internal interactions defined as [40]:

$$G_q^n(\vec{x}_1, \vec{x}_2, ..., \vec{x}_n) \equiv \langle \varphi(\vec{x}_1)...\varphi(\vec{x}_n) \rangle_q - \langle \varphi(x_1)...\varphi(x_n) \rangle_q \qquad (63)$$

The probability of the microscopic dynamical configurations is given by the general relation:

$$P(conf) = e^{-\beta S_{conf}} \qquad (64)$$

where $\beta = 1/kt$ and $S_{conf}$ is the action of the system, while the partition function $Z$ of the system is given by the relation:

$$Z = \sum_{conf} e^{-\beta S_{conf}} \qquad (65)$$

The $q$ – extension of the above statistical theory can be obtained by the $q$ – partition function $Z_q$. The $q$ – partition function is related with the meta-equilibrium distribution of the canonical ensemble which is given by the relation:

$$p_i = e_q^{-\beta q(E_i - V_q)/Z_q} \qquad (66)$$

with

$$Z_q = \sum_{conf} e_q^{-\beta q(E_i - V_q)} \qquad (67)$$

and

$$\beta_q = \beta / \sum_{conf} p_i^q \qquad (68)$$

where $\beta = 1/KT$ is the Lagrange parameter associated with the energy constraint:

$$\langle E \rangle_q \equiv \sum_{conf} p_i^q E_i / \sum_{conf} p_i^q = U_q \qquad (69)$$

The $q$ – extension of thermodynamics is related with the estimation of $q$ – Free energy ($F_q$) the $q$ – expectation value of internal energy $(U_q)$ the $q$ – specific heat $(C_q)$ by using the $q$ – partition function:

$$F_q \equiv U_q - TS_q = -\frac{1}{\beta} \ln_q Z_q \qquad (70)$$

$$U_q = \frac{\partial}{\partial \beta} \ln_q Z_q, \frac{1}{T} = \frac{\partial S_q}{\partial U_q} \qquad (71)$$

$$C_q \equiv T \frac{\partial \delta_q}{\partial T} = \frac{\partial U_q}{\partial T} = -T \frac{\partial^2 F_q}{\partial T^2} \qquad (72)$$

## 2.6.3 The Tsallis $q$ – extension of statistics via the fractal extension of dynamics.

At the equilibrium thermodynamical state the underlying statistical dynamics is Gaussian ($q=1$). As the system goes far from equilibrium the underlying statistical dynamics becomes non-Gaussian ($q \neq 1$). At the first case the phase space includes ergodic motion corresponding to normal diffusion process with mean-squared jump distances proportional to the time $\langle x^2 \rangle \sim t$ whereas far from equilibrium the phase space motion of the dynamics becomes chaotically self-organized corresponding to anomalous diffusion process with mean-squared jump distances $\langle x^2 \rangle \sim t^a$, with $a<1$ for sub-diffusion and $a>1$ for super-diffusion. The equilibrium normal-diffusion process is described by a chain equation of the Markov-type:

$$W(x_3,t_3;x_1,t_1) = \int dx_2 W(x_3,t_3;x_2,t_2) W(x_2,t_2;x_1,t_1) \qquad (73)$$

where $W(x,t;x',t')$ is the probability density for the motion from the dynamical state $(x',t')$ to the state $(x,t)$ of the phase space. The Markov process can be related to a random differential Langevin equation with additive white noise and a corresponding Fokker – Planck probabilistic equation [6] by using the initial condition:

$$\lim_{\Delta t \to 0} W(x,y;\Delta t) = \delta(x-y) \qquad (74)$$

This relation means no memory in the Markov process and help to obtain the expansion:

$$W(x,y;\Delta t) = \delta(x-y) + a(y;\Delta t)\delta'(x-y) + \frac{1}{2}b(y;\Delta t)\delta''(x-y) \qquad (75)$$

where $A(y;\Delta t)$ and $B(y;\Delta t)$ are the first and second moment of the transfer probability function $W(x,y;\Delta t)$:

$$a(y;\Delta t) = \int dx(x-y)W(x,y;\Delta t) \equiv \langle\langle \Delta y \rangle\rangle \qquad (76)$$

$$b(y;\Delta t) = \int dx(x-y)^2 W(x,y;\Delta t) \equiv \langle\langle (\Delta y)^2 \rangle\rangle \qquad (77)$$

By using the normalization condition:

$$\int dy W(x,y;\Delta t) = 1 \qquad (78)$$

we can obtain the relation:

$$a(y;\Delta t) = -\frac{1}{2}\frac{\partial b(y;\Delta t)}{\partial y} \qquad (79)$$

The Fokker – Planck equation which corresponds to the Markov process can be obtained by using the relation:

$$\frac{\partial p(x,t)}{\partial t} = \lim_{\Delta t \to 0} \frac{1}{\Delta t}\left[\int_{-\infty}^{+\infty} dy W(x,y;\Delta t)p(y,t) - p(x,t)\right] \qquad (80)$$

where $p(x,t) \equiv W(x,x_0;t)$ is the probability distribution function of the state $(x,t)$ corresponding to large time asymptotic, as follows:

$$\frac{\partial P(x,t)}{\partial t} = -\nabla_x(AP(x,t)) + \frac{1}{2}\nabla_x^2(BP(x,t)) \qquad (81)$$

where $A(x)$ is the flow coefficient:

$$A(x,t) \equiv \lim_{\Delta t \to 0} \frac{1}{\Delta t} \langle\langle \Delta x \rangle\rangle \tag{82}$$

and $B(x,t)$ is the diffusion coefficient:

$$B(\vec{x},t) \equiv \lim_{\Delta t \to 0} \frac{1}{\Delta t} \langle\langle \Delta x^2 \rangle\rangle \tag{83}$$

The Markov process is a Gaussian process as the moments $\lim_{\Delta t \to 0} \langle\langle \Delta x^m \rangle\rangle$ for $m > 2$ are zero. The stationary solutions of F-P equation satisfy the extremal condition of Boltzmann – Gibbs entropy:

$$S_{BG} = -K_B \int p(x) \ln p(x) dx \tag{84}$$

corresponding to the known Gaussian distribution:

$$p(x) \sim \exp(-x^2 / 2\sigma^2) \tag{85}$$

According to Zaslavsky [6] the fractal extension of Fokker – Planck (F-P) equation can be produced by the scale invariance principle applied for the phase space of the non-equilibrium dynamics. As it was shown by Zaslavsky for strong chaos the phase space includes self similar structures of islands inside islands dived in the stochastic sea. The fractal extension of the FPK equation (FFPK) can be derived after the application of a Renormalization group of anomalous kinetics (RGK):

$$\hat{R}_K : s' = \lambda_s S, \; t' = \lambda_t t$$

where $s$ is a spatial variable and t is the time.
Correspondingly to the Markov process equations:

$$\frac{\partial^\beta p(\xi,t)}{\partial t^\beta} \equiv \lim_{\Delta t \to 0} \frac{1}{(\Delta t)^\beta} \left[ W(\xi,\xi_0; t+\Delta t) - W(\xi,\xi_0; t) \right] \tag{86}$$

$$W(\xi, n; \Delta t) = \delta(\xi - n) + A(n; \Delta t)\delta^{(\alpha)}(\xi - n) + \frac{1}{2} B(n; \Delta t)\delta^{(2a)}(\xi - n) + \ldots \tag{87}$$

as the space-time variations of probability W are considered on fractal space-time variables $(t, \xi)$ with dimensions $(\beta, a)$.
For fractal dynamics $a(n; \Delta t)$, $b(n; \Delta t)$ satisfy the equations:

$$a(n; \Delta t) = \int |n - \xi|^\alpha W(\xi, n; \Delta t) d\xi \equiv \langle\langle |\Delta \xi|^a \rangle\rangle \tag{88}$$

$$b(n; \Delta t) = \int |n - \xi|^{2\alpha} W(\xi, n; \Delta t) d\xi \equiv \langle\langle |\Delta \xi|^{2a} \rangle\rangle \tag{89}$$

and the limit equations:

$$A(\xi) = \lim_{\Delta t \to 0} \frac{a(\xi; \Delta t)}{(\Delta t)^\beta} \tag{90}$$

$$B(\xi) = \lim_{\Delta t \to 0} \frac{b(\xi; \Delta t)}{(\Delta t)^\beta} \tag{91}$$

By them we can obtain the FFPK equation.
Far from equilibrium the non-linear dynamics can produce phase space topologies corresponding to various complex attractors of the dynamics. In this case the extended complexity of the dynamics corresponds to the generalized strange kinetic Langevin equation with correlated and multiplicative noise components and extended fractal Fokker – Planck - Kolmogorov equation (FFPK) [6, 7]. The $q$ – extension of statistics by Tsallis can be related with the strange kinetics and the fractal extension of dynamics through the Levy process:

$$P(x_n, t_n; x_0 t_0) = \int dx_1 ... dx_{N-1} P(x_N, t_N; x_{N-1}, t_{N-1}) ... P(x_1, t_1; x_0, t_0) \quad (92)$$

The Levy process can be described by the fractal F-P equation:

$$\frac{\partial^\beta P(x,t)}{\partial t^\beta} = \frac{\partial^a}{\partial (-x)^a}[A(x)P(x,t)] + \frac{\partial^{a+1}}{\partial (-x)^{a+1}}[B(x)P(x,t)] \quad (93)$$

where $\partial^\beta / \partial t^\beta$, $\partial^a / \partial(-x)^a$ and $\partial^{a+1}/\partial(-x)^{a+1}$ are the fractal time and space derivatives correspondingly [6]. The stationary solution of the F F-P equation for large $x$ is the Levy distribution $P(x) \sim x^{-(1+\gamma)}$. The Levy distribution coincides with the Tsallis $q-$extended optimum distribution q exponational function for $q = (3+\gamma)/(1+\gamma)$. The fractal extension of dynamics takes into account non-local effects caused by the topological heterogeneity and fractality of the self-organized phase – space. Also the fractal geometry and the complex topology of the phase – space introduce memory in the complex dynamics which can be manifested as creation of long range correlations, while, oppositely, in Markov process we have complete absence of memory.

In general, the fractal extension of dynamics as it was done until now from Zaslavsky, Tarasov and other scientists indicate the internal consistency of Tsallis $q-$statistics as the non-equilibrium extension of B-G statistics with the fractal extension of classical and quantum dynamics. Concerning the space plasmas the fractal character of their dynamics has been indicated also by many scientists. Indicatively, we refer the fractal properties of sunspots and their formation by fractal aggregates as it was shown by Zelenyi and Milovanov [33, 35], the anomalous diffusion and intermittent turbulence of the solar convection and photospheric motion shown by Ruzmakin et al. [36], the multi-fractal and multi-scale character of space plasmas indicated by Lui [45] and Pavlos et al. [13].

Finally we must notice the fact that the fractal extension of dynamics identifies the fractal distribution of a physical magnitude in space and time according to the scaling relation $M(R) \sim R^a$ with the fractional integration as an integration in a fractal space [9]. From this point of view it could be possible to conclude the novel concept that the non-equilibrium $q-$extension of statistics and the fractal extension of dynamics are related with the fractal space and time themselves [4, 18, 7].

## 2.6.4 Fractal acceleration and fractal energy dissipation

The problem of kinetic or magnetic energy dissipation in fluid and plasmas as well as the bursty acceleration processes of particles at flares, magnetospheric plasma sheet and other regions of space plasmas is an old and yet resisting problem of fluids or space plasma science.

Normal Gaussian diffusion process described by the Fokker – Planck equation is unable to explain either the intermittent turbulence in fluids or the bursty character of energetic particle acceleration following the bursty development of inductive electric fields after turbulent magnetic flux change in plasmas [74]. However the fractal extension of dynamics and Tsallis extension of statistics indicate the possibility for a mechanism of fractal dissipation and fractal acceleration process in fluids and plasmas.

According to Tsallis statistics and fractal dynamics the super-diffusion process:

$$\langle R^2 \rangle \sim t^\gamma \quad (94)$$

with $\gamma > 1$ ($\gamma = 1$ for normal diffusion) can be developed at systems far from equilibrium. Such process is known as intermittent turbulence or as anomalous diffusion which can be caused by Levy flight process included in fractal dynamics and fractal Fokker – Planck Kolmogorov equation (FFPK). The solution of FFPK equation [6] corresponds to double (temporal, spatial) fractal characteristic function:

$$P(k,t) = \exp\left(-const x t^{\beta} |k| a\right) \qquad (95)$$

Where $P(k,t)$ is the Fourier transform of asymptotic distribution function:

$$P(\xi,t) \sim const x t^{\beta} / \xi^{1+\alpha}, \ (\xi \to \infty) \qquad (96)$$

This distribution is scale invariant with mean displacement:

$$\langle |\xi|^{\alpha} \rangle \simeq const x t^{\beta}, \ (t \to \infty) \qquad (97)$$

According to this description, the flights of multi-scale and multi-fractal profile can explain the intermittent turbulence of fluids, the bursty character of magnetic energy dissipation and the bursty character of induced electric fields and charged particle acceleration in space plasmas as well as the non-Gaussian dynamics of brain-heart dynamics. The fractal motion of charged particles across the fractal and intermittent topologies of magnetic – electric fields is the essence of strange kinetic [6, 7]. Strange kinetics permits the development of local sources with spatial fractal – intermittent condensation of induced electric-magnetic fields in brain, heart and plasmas parallely with fractal – intermittent dissipation of magnetic field energy in plasmas and fractal acceleration of charged particles. Such kinds of strange accelerators in plasmas can be understood by using the Zaslavsky studies for Hamiltonian chaos in anomalous multi-fractal and multi-scale topologies of phase space [6]. Generally the anomalous topology of phase space and fractional Hamiltonian dynamics correspond to dissipative non-Hamiltonian dynamics in the usual phase space [9]. The most important character of fractal kinetics is the wandering of the dynamical state through the gaps of cantori which creates effective barriers for diffusion and long range Levy flights in trapping regions of the phase space. Similar Levy flights processes can be developed by the fractal dynamics and intermittent turbulence of the complex systems.

In this theoretical framework it is expected the existence of Tsallis non extensive entropy and q-statistics in non-equilibrium distributed complex systems as, fluids, plasmas or brain and heart systems which are studied in the next section of this work. The fractal dynamics corresponding to the non-extensive Tsallis $q$ – statistical character of the probability distributions in the distributed complex systems indicate the development of a self-organized and globally correlated parts of active regions in the distributed dynamics. This character can be related also with deterministic low dimensional chaotic profile of the active regions according to Pavlos et al. [13].

## 3. Theoretical expectations through Tsallis statistical theory and fractal dynamics.

Tsallis $q$ – statistics as well as the non-equilibrium fractal dynamics indicate the multi-scale, multi-fractal chaotic and holistic dynamics of space plasmas. Before we present experimental verification of the theoretical concepts described in previous studies as concerns space plasmas in this section we summarize the most significant theoretical expectations.

## 3.1 The $q$ – triplet of Tsallis.

The non-extensive statistical theory is based mathematically on the nonlinear equation:
$$\frac{dy}{dx} = y^q, \; (y(0)=1, q \in \Re) \quad (98)$$
with solution the $q$ – exponential function defined previously in equation (2.2). The solution of this equation can be realized in three distinct ways included in the $q$ – triplet of Tsallis: ($q_{sen}, q_{stat}, q_{rel}$). These quantities characterize three physical processes which are summarized here, while the $q$ – triplet values characterize the attractor set of the dynamics in the phase space of the dynamics and they can change when the dynamics of the system is attracted to another attractor set of the phase space. The equation (2.36) for $q=1$ corresponds to the case of equilibrium Gaussian Boltzmann-Gibbs (BG) world [1]. In this case of equilibrium BG world the $q$ – triplet of Tsallis is simplified to ($q_{sen}=1, q_{stat}=1, q_{rel}=1$).

### a. The $q_{stat}$ index and the non-extensive physical states.

According to [1] the long range correlated metaequilibrium non-extensive physical process can be described by the nonlinear differential equation:
$$\frac{d(p_i Z_{stat})}{dE_i} = -\beta q_{stat}(p_i Z_{stat})^{q_{stat}} \quad (99)$$
The solution of this equation corresponds to the probability distribution:
$$p_i = e_{q_{stat}}^{-\beta_{stat} E_i} / Z_{q_{stat}} \quad (100)$$

where $\beta_{q_{stat}} = \frac{1}{KT_{stat}}$, $Z_{stat} = \sum_j e_{q_{stat}}^{-\beta q_{stat} E_j}$.

Then the probability distribution function is given by the relations:
$$p_i \propto \left[1-(1-q)\beta_{q_{stat}} E_i\right]^{1/1-q_{stat}} \quad (101)$$
for discrete energy states $\{E_i\}$ by the relation:
$$p(x) \propto \left[1-(1-q)\beta_{q_{stat}} x^2\right]^{1/1-q_{stat}} \quad (102)$$
for continuous $X$ states $\{X\}$, where the values of the magnitude $X$ correspond to the state points of the phase space.

The above distributions functions (101,102) correspond to the attracting stationary solution of the extended (anomalous) diffusion equation related with the nonlinear dynamics of system. The stationary solutions $P(x)$ describe the probabilistic character of the dynamics on the attractor set of the phase space. The non-equilibrium dynamics can be evolved on distinct attractor sets depending upon the control parameters values, while the $q_{stat}$ exponent can change as the attractor set of the dynamics changes.

### b. The $q_{sen}$ index and the entropy production process

The entropy production process is related to the general profile of the attractor set of the dynamics. The profile of the attractor can be described by its multifractality as well as by its sensitivity to initial conditions. The sensitivity to initial conditions can be described as follows:

$$\frac{d\xi}{d\tau} = \lambda_1 \xi + (\lambda_q - \lambda_1)\xi^q \tag{103}$$

where $\xi$ describes the deviation of trajectories in the phase space by the relation: $\xi \equiv \lim_{\Delta(x) \to 0} \{\Delta x(t) \backslash \Delta x(0)\}$ and $\Delta x(t)$ is the distance of neighboring trajectories. The solution of equation (2.41) is given by:

$$\xi = \left[1 - \frac{\lambda q_{sen}}{\lambda_1} + \frac{\lambda q_{sen}}{\lambda_1} e^{(1-q_{sen})\lambda_1 t}\right]^{\frac{1}{1-q}} \tag{104}$$

The $q_{sen}$ exponent can be also related with the multifractal profile of the attractor set by the relation:

$$\frac{1}{q_{sen}} = \frac{1}{a_{min}} - \frac{1}{a_{max}} \tag{105}$$

where $a_{min}(a_{max})$ corresponds to the zero points of the multifractal exponent spectrum $f(a)$. That is $f(a_{min}) = f(a_{max}) = 0$.

The deviations of neighboring trajectories as well as the multifractal character of the dynamical attractor set in the system phase space are related to the chaotic phenomenon of entropy production according to Kolmogorov – Sinai entropy production theory and the Pesin theorem. The $q$ – entropy production is summarized in the equation:

$$K_q \equiv \lim_{t \to \infty} \lim_{W \to \infty} \lim_{N \to \infty} \frac{<S_q>(t)}{t}. \tag{106}$$

The entropy production ($dS_q / t$) is identified with $K_q$, as $W$ are the number of non-overlapping little windows in phase space and $N$ the state points in the windows according to the relation $\sum_{i=1}^{W} N_i = N$. The $S_q$ entropy is estimated by the probabilities $P_i(t) \equiv N_i(t)/N$. According to Tsallis the entropy production $K_q$ is finite only for $q = q_{sen}$.

**c. The $q_{rel}$ index and the relaxation process**

The thermodynamical fluctuation – dissipation theory is based on the Einstein original diffusion theory (Brownian motion theory). Diffusion process is the physical mechanism for extremization of entropy. If $\Delta S$ denote the deviation of entropy from its equilibrium value $S_0$, then the probability of the proposed fluctuation that may occur is given by:

$$P \sim \exp(\Delta s / k). \tag{107}$$

The Einstein – Smoluchowski theory of Brownian motion was extended to the general Fokker – Planck diffusion theory of non-equilibrium processes. The potential of Fokker – Planck equation may include many metaequilibrium stationary states near or far away from the basic thermodynamical equilibrium state. Macroscopically, the relaxation to the equilibrium stationary state can be described by the form of general equation as follows:

$$\frac{d\Omega}{d\tau} \simeq -\frac{1}{\tau}\Omega, \tag{108}$$

where $\Omega(t) \equiv [O(t) - O(\infty)]/[O(0) - O(\infty)]$ describes the relaxation of the macroscopic observable $O(t)$ relaxing towards its stationary state value. The non-extensive generalization of fluctuation – dissipation theory is related to the general correlated anomalous diffusion processes. Now, the equilibrium relaxation process (101) is transformed to the metaequilibrium non-extensive relaxation process:

$$\frac{d\Omega}{dt} = -\frac{1}{T_{q_{rel}}}\Omega^{q_{rel}} \qquad (109)$$

the solution of this equation is given by:

$$\Omega(t) \simeq e_{q_{rel}}^{-t/\tau_{rel}} \qquad (110)$$

The autocorrelation function $C(t)$ or the mutual information $I(t)$ can be used as candidate observables $\Omega(t)$ for the estimation of $q_{rel}$. However, in contrast to the linear profile of the correlation function, the mutual information includes the non linearity of the underlying dynamics and it is proposed as a more faithful index of the relaxation process and the estimation of the Tsallis exponent $q_{rel}$.

### 3.2 Measures of Multifractal Intermittence Turbulence

In the following, we follow Arimitsu and Arimitsu [72] for the theoretical estimation of significant quantitative relations which can also be estimated experimentally. The probability singularity distribution $P(a)$ can be estimated as extremizing the Tsallis entropy functional $S_q$. According to Arimitsu and Arimitsu [72] the extremizing probability density function $P(a)$ is given as a $q$-exponential function:

$$P(a) = Z_q^{-1}[1-(1-q)\frac{(a-a_0)^2}{2X/\ln 2}]^{\frac{1}{1-q}} \qquad (111)$$

where the partition function $Z_q$ is given by the relation:

$$Z_q = \sqrt{2X/[(1-q)\ln 2]}\ B(1/2, 2/1-q), \qquad (112)$$

and $B(a,b)$ is the Beta function. The partition function $Z_q$ as well as the quantities $X$ and $q$ can be estimated by using the following equations:

$$\left.\begin{array}{l}\sqrt{2X} = \left[\sqrt{a_0^2 + (1-q)^2} - (1-q)\right]/\sqrt{b} \\ b = (1-2^{-(1-q)})/[(1-q)\ln_2]\end{array}\right\} \qquad (113)$$

We can conclude for the exponent's spectrum $f(a)$ by using the relation $P(a) \approx \ln^{d-F(a)}$ as follows:

$$f(a) = D_0 + \log_2[1-(1-q)\frac{(a-a_o)^2}{2X/\ln 2}]/(1-q)^{-1} \qquad (114)$$

where $a_0$ corresponds to the $q$-expectation (mean) value of $a$ through the relation:

$$<(a-a_0)^2>_q = (\int da P(a)^q (a-a_0)^q)/\int da P(a)^q. \qquad (115)$$

while the $q$-expectation value $a_0$ corresponds to the maximum of the function $f(a)$ as $df(a)/da|_{a_0} = 0$. For the Gaussian dynamics $(q \to 1)$ we have mono-fractal spectrum $f(a_0) = D_0$. The mass exponent $\tau(\bar{q})$ can be also estimated by using the inverse Legendre transformation: $\tau(\bar{q}) = a\bar{q} - f(a)$ (relations 26-27) and the relation (114) as follows:

$$\tau(\bar{q}) = \bar{q}a_0 - 1 - \frac{2X\bar{q}^2}{1+\sqrt{C_{\bar{q}}}} - \frac{1}{1-q}[1-\log_2(1+\sqrt{C_{\bar{q}}})], \qquad (116)$$

Where $C_{\bar{q}} = 1 + 2\bar{q}^2(1-q)X \ln 2$.

The relation between $a$ and $q$ can be found by solving the Legendre transformation equation $\bar{q} = df(a)/da$. Also if we use the equation (114) we can obtain the relation:

$$a_{\bar{q}} - a_0 = (1 - \sqrt{C_{\bar{q}}})/[\bar{q}(1-q)\ln 2] \qquad (117)$$

The $q$-index is related to the scaling transformations (22) of the multifractal nature of turbulence according to the relation $q = 1 - a$. Arimitsu and Arimitsu [72] estimated the $q$-index by analyzing the fully developed turbulence state in terms of Tsallis statistics as follows:

$$\frac{1}{1-q} = \frac{1}{a_-} - \frac{1}{a_+} \qquad (118)$$

where $a_\pm$ satisfy the equation $f(a_\pm) = 0$ of the multifractal exponents spectrum $f(a)$. This relation can be used for the estimation of $q_{sen}$-index included in the Tsallis $q$-triplet (see next section).

## 4. Comparison of theory with the observations

### 4.1 The Tsallis q-statistics

After this general investment in the following, we present evidence of Tsallis non-extensive $q$-statistics for space plasmas. The Tsallis statistics in relation with fractal and chaotic dynamics of space plasmas will be presented in a short coming series of publications.

In next sections we present estimations of Tsallis statistics for various kinds of space plasma's systems. The $q_{stat}$ Tsallis index was estimated by using the observed Probability Distribution Functions (PDF) according to the Tsallis q-exponential distribution:

$$PDF[\Delta Z] \equiv A_q \left[1 + (q-1)\beta_q (\Delta Z)^2\right]^{\frac{1}{1-q}}, \qquad (119)$$

where the coefficients $A_q$, $\beta_q$ denote the normalization constants and $q \equiv q_{stat}$ is the entropic or non-extensivity factor ($q_{stat} \leq 3$) related to the size of the tail in the distributions. Our statistical analysis is based on the algorithm described in [75]. We construct the $PDF[\Delta Z]$ which is associated to the first difference $\Delta Z = Z_{n+1} - Z_n$ of the experimental sunspot time series, while the $\Delta Z$ range is subdivided into little ``cells'' (data binning process) of width $\delta z$, centered at $z_i$ so that one can assess the frequency of $\Delta z$-values that fall within each cell/bin. The selection of the cell-size $\delta z$ is a crucial step of the algorithmic process and its equivalent to solving the binning problem: a proper initialization of the bins/cells can speed up the statistical analysis of the data set and lead to a convergence of the algorithmic process towards the exact solution. The resultant histogram is being properly normalized and the estimated q-value corresponds to the best linear fitting to the graph $\ln_q(p(z_i))$ vs $z_i^2$. Our algorithm estimates for each $\delta_q = 0,01$ step the linear adjustment on the graph under scrutiny (in this case the $\ln_q(p(z_i))$ vs $z_i^2$ graph) by evaluating the associated correlation coefficient *(CC)*, while the best linear fit is considered to be the one maximizing the correlation coefficient. The obtained $q_{stat}$, corresponding to the best linear adjustment is then being used to compute the following equation:

$$G_q(\beta, z) = \frac{\sqrt{\beta}}{C_q} e_q^{-\beta z^2} \qquad (120)$$

where $C_q = \sqrt{\pi} \cdot \Gamma(\frac{3-q}{2(q-1)}) / \sqrt{q-1} \cdot \Gamma(\frac{1}{q-1})$, $1 < q < 3$ for different $\beta$-values. Moreover, we select the $\beta$-value minimizing the $\sum_i [G_{q_{sstat}}(\beta, z_i) - p(z_i)]^2$, as proposed again in [75].

In the following we present the estimation of Tsallis statistics $q_{stat}$ for various cases of space plasma system. Especially, we study the $q-$ statistics for the following space plasma complex systems: I Magnetospheric system, II Solar Wind (magnetic cloud), III Solar activity, IV Cosmic stars, IIV Cosmic Rays.

**4.2 Cardiac Dynamics**

For the study of the q-triplet statistics we used measurements from the cardiac and especially the heart rate variability timeseries which includes a multivariate data set recorded from a patient in the sleep laboratory of the Beth Israel Hospital in Boston, Massachusetts. The heart rate was determined by measuring the time between the QRS complexes in the electrocardiogram, taking the inverse, and then converting this to an evenly sampled record by interpolation. They were converted from 250 Hz to 2 Hz data by averaging over a 0.08 second window at the times of the heart rate samples.

In Fig.1[a] we present the timeseries of hrv and in Fig.1[b] (by open circles) we present the experimental probability distribution function (PDF) p(z) vs. z, where z corresponds to the $Z_{n+1} - Z_n, (n = 1, 2, ..., N)$ timeseries difference values. In Fig.1[c] we present the best linear correlation between $\ln_q[p(z)]$ and $z^2$. The best fitting was found for the value of $q_{stat} = 1.26 \pm 0.10$.

Fig.1[d-e] presents the estimation of the generalized dimension $D_q$ and their corresponding multifractal (or singularity) spectrum $f(\alpha)$, from which the $q_{sen}$ index was estimated by using the relation $1/(1-q_{sens}) = 1/a_{min} - 1/a_{max}$. In Fig.1[d] the experimentally estimated spectrum function $f(\alpha)$ is compared with a polynomial of fourth order (red line) as well as by the theoretically estimated function $f(\alpha)$ (green line), by using the Tsallis $q-$ entropy. As we can observe the theoretical estimation is faithful with high precision on the whole part of the experimental function $f(a)$.

Similar comparison of the theoretical prediction and the experimental estimation of the generalized dimensions function $D(q)$ is shown in Fig.1[e]. In these figures the solid red line correspond to the $p$-model prediction, while the solid green line correspond to the $D(q)$ function estimation according to Tsallis theory. The correlation coefficient of the fitting was found higher than $0.9$.

Fig.1[f] presents the best log plot fitting of the autocorrelation function $C(\tau)$ estimated for the hrv data set. The q-triplet values were found to satisfy the relation $q_{rel} < 1 < q_{stat} < q_{rel}$:
($-0.77 \pm 0.06 < 1 < 1.26 \pm 0.10 < 3.32 \pm 0.44$).

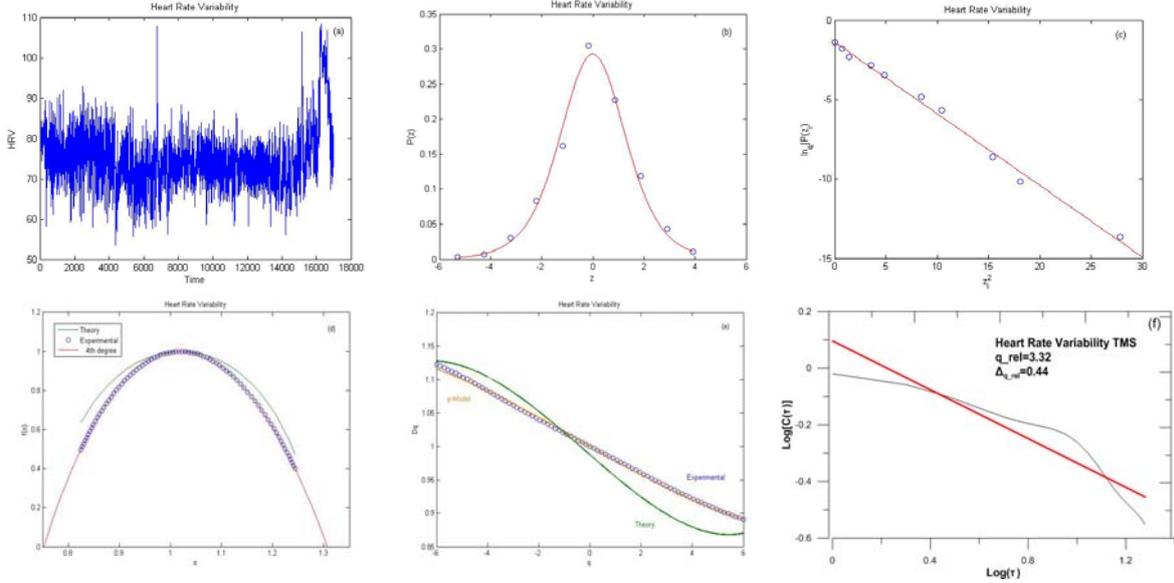

*Figure 1: (a) Time series of heart rate variability (b) PDF P(z_i) vs. z_i q Guassian function that fits P(z_i) for the heart rate variability (c) Linear Correlation between ln_qP(z_i) and (z_i)² where q_stat = 1.26 ± 0.10 for the heart rate variability (d) Multifractal spectrum of hear rate variability time series with solid line a fourth degree polynomial. We calculate the q_sen = -0.772 ± 0.068. (e) D(q) vs. q of heart rate variability time series (f) Log – log plot of the self-correlation coefficient C(τ) vs. time delay τ for the heart rate variability time series. We obtain the best fit with q_rel = 3.32 ± 0.44.*

### 4.3 Brain Dynamics

For the study of the q-triplet statistics we used measurements from real EEG timeseries from epileptic patients during seizure attack. Each EEG timeseries consisting of 3.750 points. The width of the timeseries is ranging from -1,000 Volt to 1,000 Volt.

In Fig.2[a] we present the timeseries of EEG timeseries (seizure state) and in Fig.1[b] (by open circles) we present the experimental probability distribution function (PDF) p(z) vs. z. In Fig.2[c] we present the best linear correlation between $\ln_q[p(z)]$ and $z^2$. The best fitting was found for the value of $q_{stat} = 1.63 \pm 0.14$.

Fig.2[d-e] presents the estimation of the generalized dimension $D_q$ and their corresponding multifractal (or singularity) spectrum $f(\alpha)$, from which the $q_{sen}$ index was estimated by using

the relation $1/(1-q_{sens}) = 1/a_{min} - 1/a_{max}$. In Fig.2[d] the experimentally estimated spectrum function $f(\alpha)$ is compared with a polynomial of sixth order (red line) as well as by the theoretically estimated function $f(\alpha)$ (green line), by using the Tsallis $q-$entropy. As we can observe the theoretical estimation is faithful with high precision on the first part of the experimental function $f(a)$. Similar comparison of the theoretical prediction and the experimental estimation of the generalized dimensions function $D(q)$ is shown in Fig.2[e]. In these figures the solid red line correspond to the $p-$model prediction, while the solid green line correspond to the $D(q)$ function estimation according to Tsallis theory. The correlation coefficient of the fitting was found higher than $0.9$.

Fig.2[f] presents the best log plot fitting of the autocorrelation function $C(\tau)$ estimated for the EEG data set. The q-triplet values were found to satisfy the relation $q_{rel} < 1 < q_{stat} < q_{rel}$: ($-0.451 \pm 0.036 < 1 < 1.63 \pm 0.14 < 10.43 \pm 0.95$).

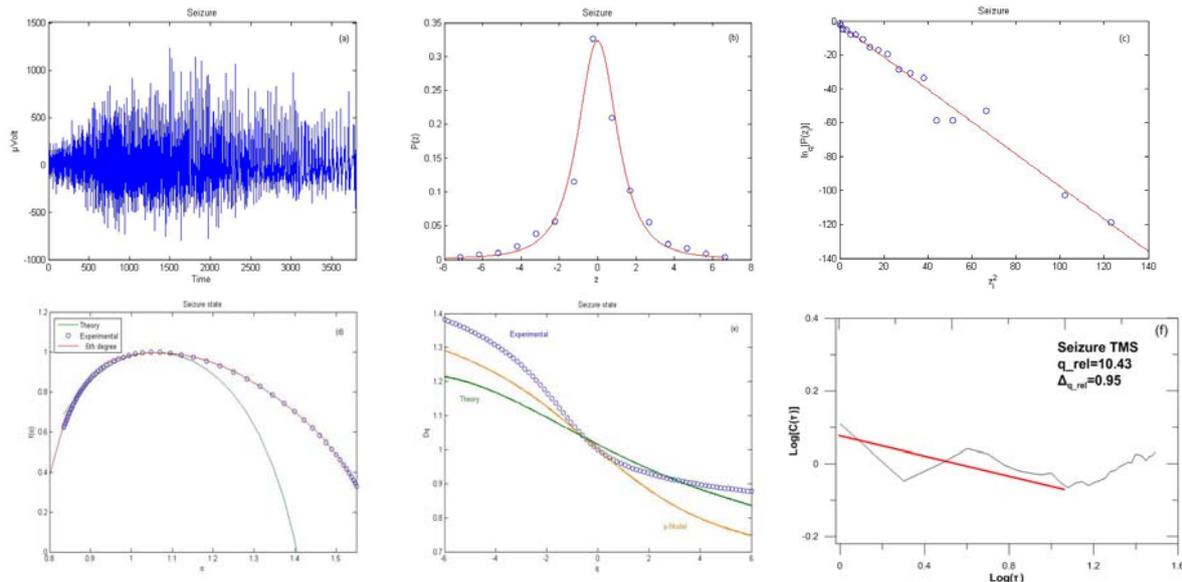

*Figure 2: (a)* Time series of seizure state *(b)* PDF $P(z_i)$ vs. $z_i$ q Guassian function that fits $P(z_i)$ for the seizure state *(c)* Linear Correlation between $ln_q P(z_i)$ and $(z_i)^2$ where $q = 1.63 \pm 0.14$ for the seizure state *(d)* Multifractal spectrum of seizure state t time series with solid line a sixth degree polynomial. We calculate the $q_{sen} = -0.451 \pm 0.036$. *(e)* $D(q)$ vs. q of the seizure state time series *(f)* Log – log plot of the self-correlation coefficient $C(\tau)$ vs. time delay $\tau$ for the seizure state time series. We obtain the best fit with $q_{rel} = 10.43 \pm 0.95$.

## 4.4 Earthquake Dynamics

In this sub-section we present the q-triplet Tsallis statistics of the experimental data from earthquakes in the region of whole Greece with magnitude greater from 4 and time period 1964-2004. The data set was found from the National Observatory of Athens (NOA).

In Figure 3a the time series of Interevent Times is presented, while the corresponding q-value is shown in Figure 3b and was found to be $q_{stat} = 2.28 \pm 0.12$. In Figure 3g we present the experimental time series of Magnitude data. The q-statistics for this case are presented in Figure 3h. The corresponding q-value was found to be $q_{stat} = 1.77 \pm 0.09$. The results reveal clearly non-Gaussian statistics for the earthquake Interevent Times and Magnitude data. The results showed the existence of q-statistics and the non-Gaussianity of the data sets.

Moreover in Fig.[d-e] and Fig.[j-k] we showed the multifractal (singularity) spectrum $f(\alpha)$ and the generalized dimensions function $D(q)$ for the interevent times timeseries and the magnitude timeseries correspondingly. Fig.3 [f] and Fig. 3[l] presents the best log plot fitting of the autocorrelation function $C(\tau)$ estimated for the interevent times timeseries and the magnitude timeseries correspondingly.

The q-triplet values were found to satisfy the relation $q_{rel} < 1 < q_{stat} < q_{rel}$ for the two data sets from earthquakes dynamics (see table 1).

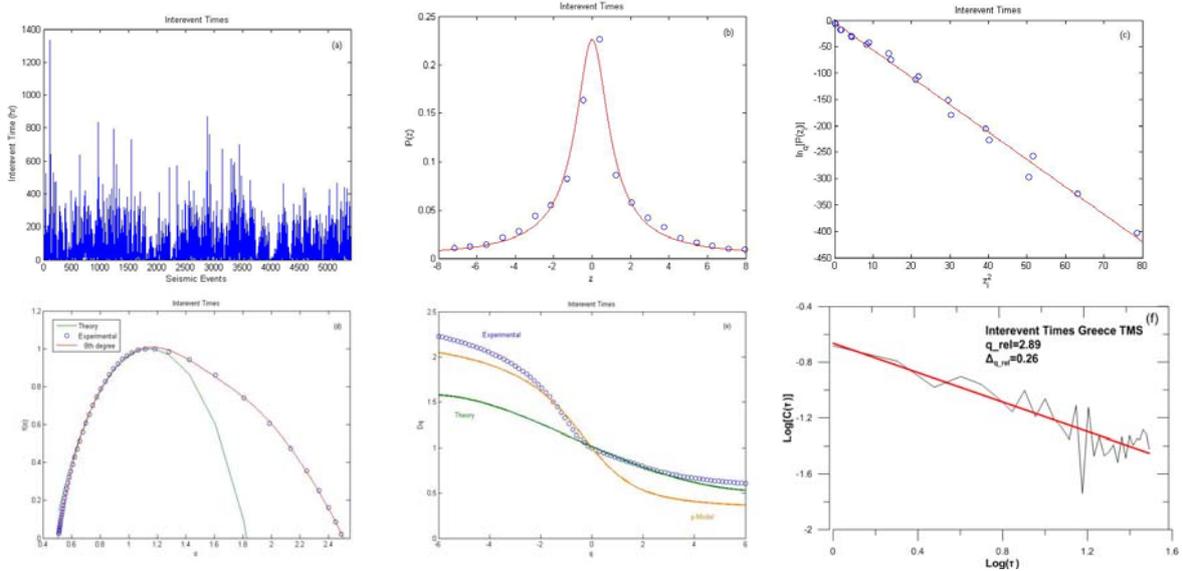

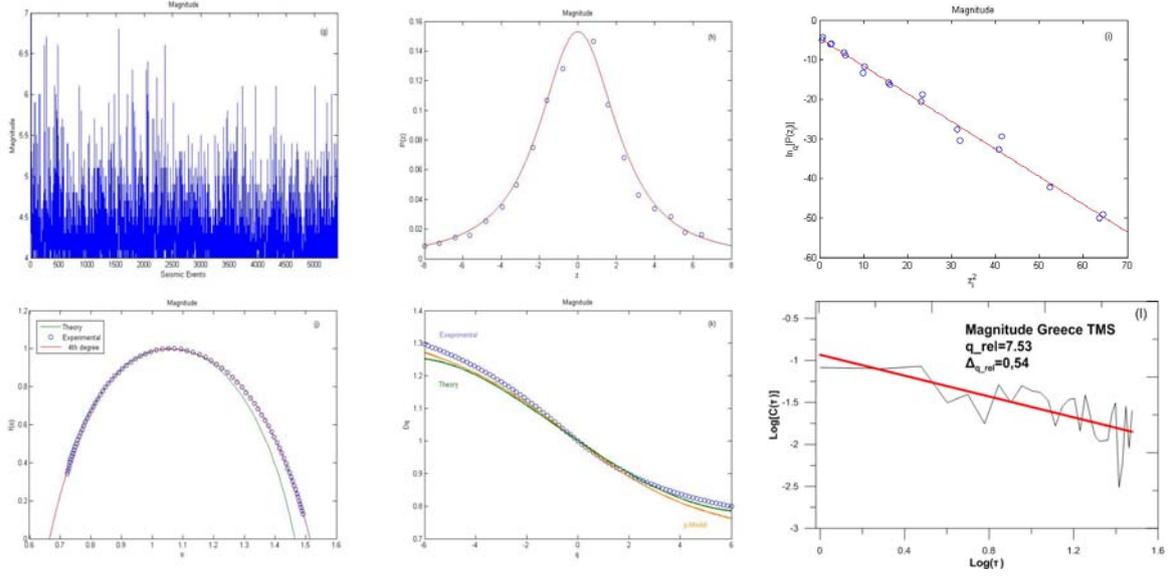

***Figure 3:*** *(a) Time series of Interevent Times (b) PDF $P(z_i)$ vs. $z_i$ q Guassian function that fits $P(z_i)$ for the Interevent Times (c) Linear Correlation between $ln_qP(z_i)$ and $(z_i)^2$ where $q_{stat} = 2.28 \pm 0.12$ for the Interevent Times (d) Multifractal spectrum of Interevent Times time series with solid line a eight degree polynomial. We calculate the $q_{sen} = 0.370 \pm 0.005$. (e) $D(q)$ vs. $q$ of the Interevent Times time series (f) Log – log plot of the self-correlation coefficient $C(\tau)$ vs. time delay $\tau$ for the Interevent Times time series. We obtain the best fit with $q_{rel} = 2.89 \pm 0.26$ (g) Time series of Magnitude (h) PDF $P(z_i)$ vs. $z_i$ q Gaussian function that fits $P(z_i)$ for the Magnitude (i) Linear Correlation between $ln_qP(z_i)$ and $(z_i)^2$ where $q_{stat} = 1.77 \pm 0.09$ for the Magnitude (j) Multifractal spectrum of Magnitude time series with solid line a fourth degree polynomial. We calculate the $q_{sen} = -0.183 \pm 0.037$. (k) $D(q)$ vs. $q$ of the Magnitude time series (l) Log – log plot of the self-correlation coefficient $C(\tau)$ vs. time delay $\tau$ for the Magnitude time series. We obtain the best fit with $q_{rel} = 7.53 \pm 0.54$.*

### 4.5 Atmospheric Dynamics

In this sub-section we study the q-triplet Tsallis statistics for the air temperature and rain fall experimental data sets from the weather station 20046 Polar GMO in E.T. Krenkelja for the period 1/1/1960 – 31/12/1960.

In Fig.4[a-f] Fig.4[g-l] we presented the estimation of the q-triplet of the experimental time series from temperature and rainfall correspondingly. In both cases we observed clearly non Gaussian statistics. The q-triplet values were found to satisfy the relation $q_{rel} < 1 < q_{stat} < q_{rel}$ for the two data sets from atmospheric dynamics (see table 1).

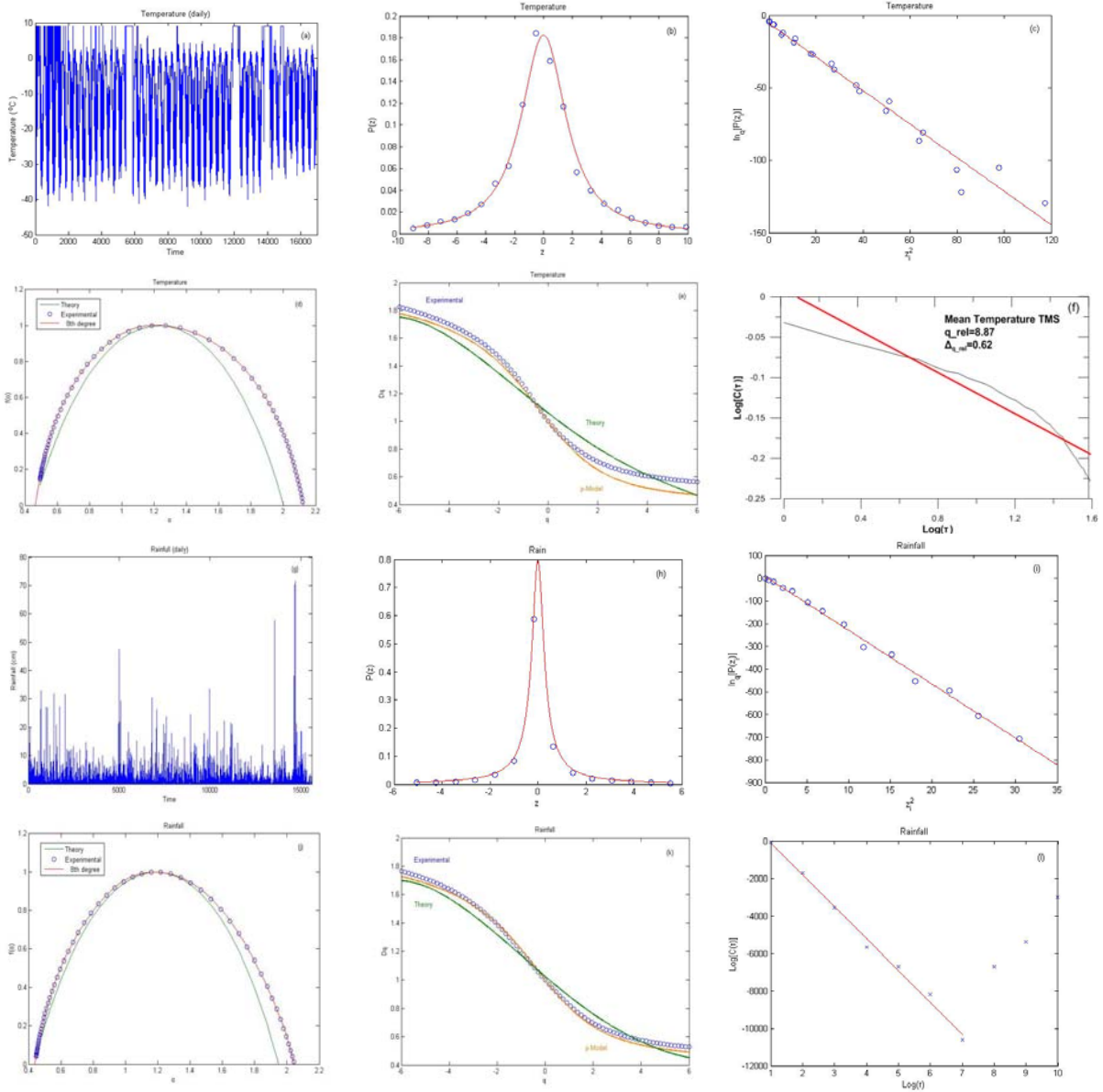

*Figure 4:* *(a)* Time series of Temperature *(b)* PDF $P(z_i)$ vs. $z_i$ q Guassian function that fits $P(z_i)$ for the Temperature *(c)* Linear Correlation between $ln_qP(z_i)$ and $(z_i)^2$ where $q_{stat} = 1.89 \pm 0.08$ for the Temperature *(d)* Multifractal spectrum of Temperature time series with solid line a eight degree polynomial. We calculate the $q_{sen} = 0.407 \pm 0.013$. *(e)* $D(q)$ vs. $q$ of the Temperature time series *(f)* Log – log plot of the self-correlation coefficient $C(\tau)$ vs. time delay $\tau$ for the Temperature time series. We obtain the best fit with $q_{rel} = 8.87 \pm 0.62$ *(g)* Time series of Rainfall *(h)* PDF $P(z_i)$ vs. $z_i$ q Gaussian function that fits $P(z_i)$ for the Rainfall *(i)* Linear Correlation between $ln_qP(z_i)$ and $(z_i)^2$ where $q_{stat} = 2.21 \pm 0.06$ for the Rainfall *(j)* Multifractal spectrum of Rainfall time series with solid line a eigth degree polynomial. We calculate the $q_{sen} = 0.444 \pm 0.007$. *(k)* $D(q)$ vs. $q$ of the Rainfall time series *(l)* Log – log plot of the self-correlation coefficient $C(\tau)$ vs. time delay $\tau$ for the Rainfall time series. We obtain the best fit with $q_{rel} = 6.04 \pm 0.47$.

### 4.6 Magnetospheric MHD Dynamics

The estimation of $B_z, V_x$ Tsallis q-triplet statistics during the substorm period is presented in Fig.5(a-l). Fig.5(a,g) shows the experimental time series corresponding to spacecraft observations of magnetic field $B_z$ and bulk plasma flows $V_x$ component. The q-values of the signals under scrutiny were found to be $q_{stat} = 1.98 \pm 0.06$ for the $V_x$ plasma velocity time series and $q_{stat} = 2.05 \pm 0.04$ for the magnetic field $B_z$ component. The fact that the magnetic field and plasma flow observations obey to non-extensive Tsallis with $q$ – values much higher than the Gaussian case ($q=1$) permit to conclude for magnetospheric plasma the existence of non-equilibrium MHD anomalous diffusion process. Similar with previous physical systems in Fig.5[d-e] and Fig.5[j-k] we showed the multifractal (singularity) spectrum $f(\alpha)$ and the generalized dimensions function $D(q)$ for the $B_z, V_x$ timeseries correspondingly

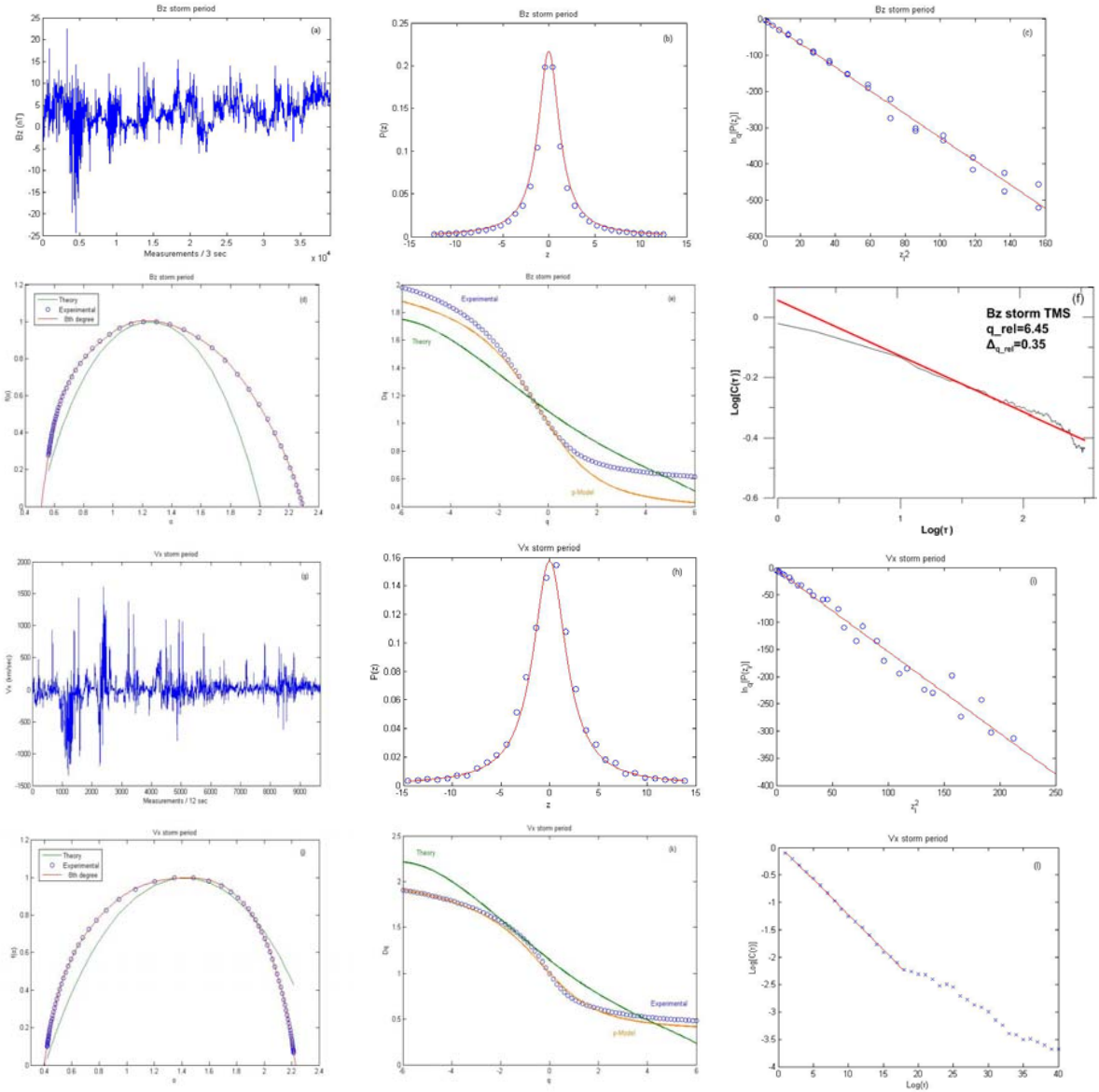

***Figure 5:*** *(a) Time series of Bz storm period (b) PDF P($z_i$) vs. $z_i$ q Gaussian function that fits P($z_i$) for the Bz storm period (c) Linear Correlation between $ln_q P(z_i)$ and $(z_i)^2$ where $q_{stat} = 2.05 \pm 0.04$ for the Bz storm period (d) Multifractal spectrum of Bz storm period time series with solid line a eight degree polynomial. We calculate the $q_{sen} = 0.341 \pm 0.033$. (e) D(q) vs. q of the Bz storm period time series (f) Log – log plot of the self-correlation coefficient C(τ) vs. time delay τ for the Bz storm period time series. We obtain the best fit with $q_{rel} = 6.45 \pm 0.35$ (g) Time series of Vx storm period (h) PDF P($z_i$) vs. $z_i$ q Guassian function that fits P($z_i$) for the Vx storm period (i) Linear Correlation between $ln_q P(z_i)$ and $(z_i)^2$ where $q_{stat} = 1.98 \pm 0.06$ for the Vx storm period (j) Multifractal spectrum of Vx storm period time series with solid line a eight degree polynomial. We calculate the $q_{sen} = 0.516 \pm 0.010$. (k) D(q) vs. q of the Vx storm period time series (l) Log – log plot of self-correlation coefficient C(τ) vs. time delay τ for the Vx storm period time series. We obtain the best fit with $q_{rel} = 2.30 \pm 0.1$*

## 4.7 Magnetospheric Charged Particles Dynamics

In the following we study the q-triplet Tsallis statistics of magnetospheric energetic particle during a strong sub-storm period. We used the data set from the GEOTAIL/EPIC experiment during the period from 12:00 UT to 21:00 UT of 8/2/1997 and from 12:00 UT of 9/2/1997 to 12:00 UT of 10/2/1997.

The Tsallis q-triplet estimated for the magnetospheric electric field and the magnetospheric particles $(e^-, p^+)$ during the storm period is shown in Fig.6[a-r]. Fig. 6[a,g,m] present the spacecraft observations of the magnetospheric electric field $E_y$ component and the magnetospheric electrons $(e^-)$ and protons $(p^+)$. The estimating Tsallis q-triplet statistics reveal clearly non-Gaussian dynamics for the mechanism of electric field development and electrons-protons acceleration during the magnetospheric storm period. Moreover, the q-triplet values were found to satisfy the relation $q_{rel} < 1 < q_{stat} < q_{rel}$ for the three data sets from Magnetospheric charged particles dynamics (see table 1).

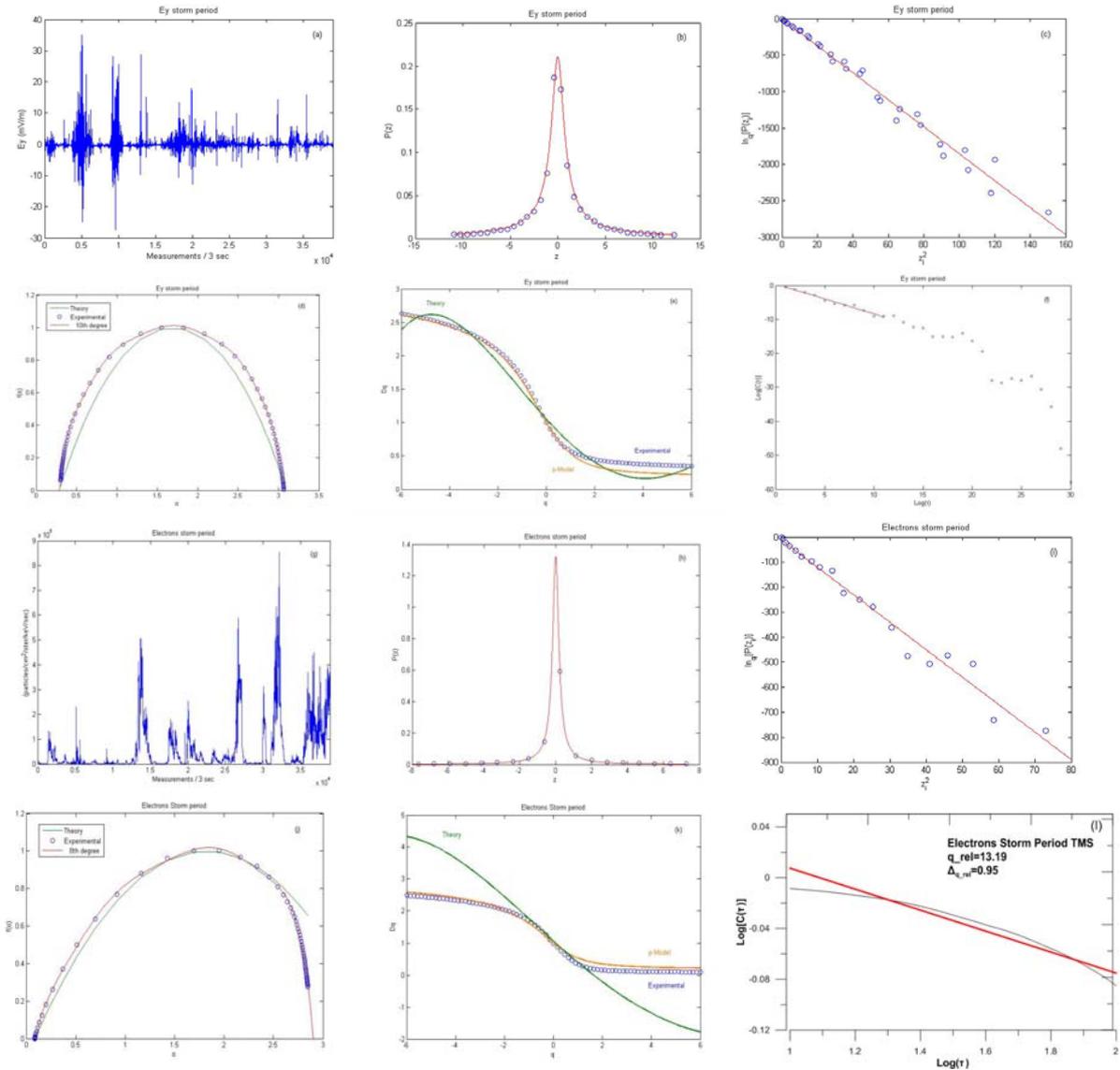

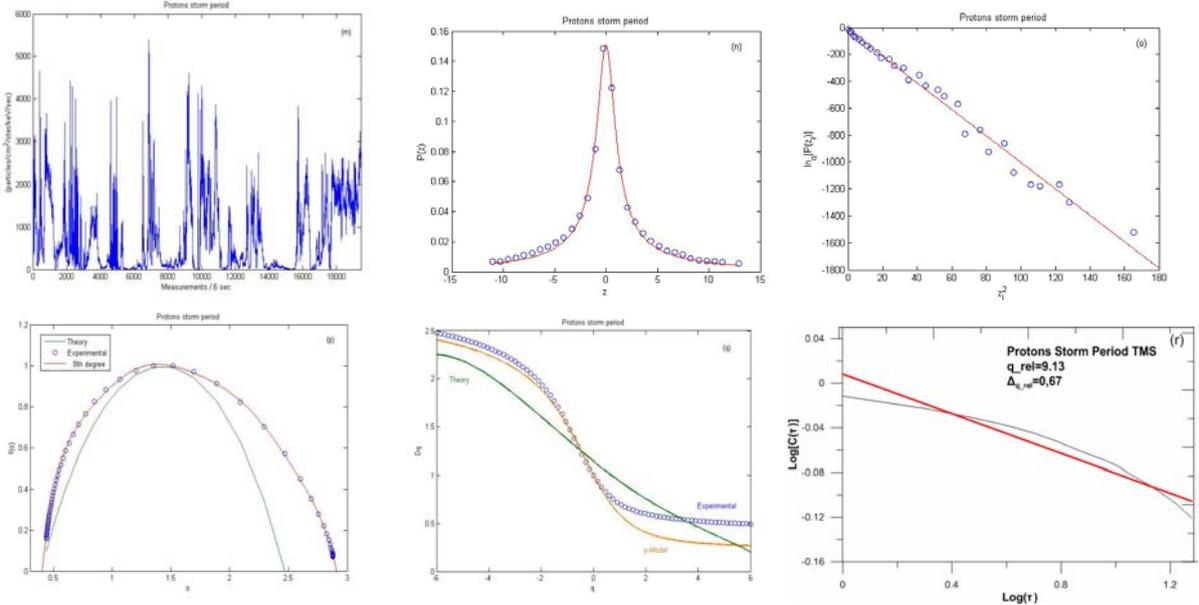

*Figure 6: (a)* Time series of Ey storm period *(b)* PDF $P(z_i)$ vs. $z_i$ q Guassian function that fits $P(z_i)$ for the Ey storm period *(c)* Linear Correlation between $ln_q P(z_i)$ and $(z_i)^2$ where $q_{stat} = 2.49 \pm 0.07$ for the Ey storm period *d)* Multifractal spectrum of Ey storm period time series with solid line a tenth degree polynomial. We calculate the $q_{sen} = 0.685 \pm 0.016$. *(e)* $D(q)$ vs. q of the Ey storm period time series *(f)* Log – log plot of the self-correlation coefficient $C(\tau)$ vs. time delay $\tau$ for the Ey storm period time series. We obtain the best fit with $q_{rel} = 3.95 \pm 0.34$ *(g)* Time series of electrons storm period *(h)* PDF $P(z_i)$ vs. $z_i$ q Gaussian function that fits $P(z_i)$ for the electrons storm period *(i)* Linear Correlation between $ln_q P(z_i)$ and $(z_i)^2$ where $q_{stat} = 2.15 \pm 0.07$ for the electrons storm period time series *(j)* Multifractal spectrum of electrons storm period time series with solid line a eigth degree polynomial. We calculate the $q_{sen} = 0.912 \pm 0.001$. *(k)* $D(q)$ vs. q of the electrons storm period time series *(l)* Log – log plot of the self-correlation coefficient $C(\tau)$ vs. time delay $\tau$ for the electrons storm period time series. We obtain the best fit with $q_{rel} = 13.19 \pm 0.95$ *(m)* Time series of protons storm period *(n)* PDF $P(z_i)$ vs. $z_i$ q Gaussian function that fits $P(z_i)$ for the protons storm period *(o)* Linear Correlation between $ln_q P(z_i)$ and $(z_i)^2$ where $q_{stat} = 2.49 \pm 0.05$ for the protons storm period. *(p)* Multifractal spectrum of protons storm period time series with solid line a eighth degree polynomial. We calculate the $q_{sen} = 0.533 \pm 0.017$. *(q)* $D(q)$ vs. q of the protons storm period time series *(r)* Log – log plot of the self-correlation coefficient $C(\tau)$ vs. time delay $\tau$ for the protons storm period time series. We obtain the best fit with $q_{rel} = 9.13 \pm 0.67$.

## 4.8 Solar Wind Magnetic Cloud Dynamics

From the spacecraft ACE, magnetic field experiment (MAG) we take raw data and focus on the Bz magnetic field component with a sampling rate 3 sec. Tha data correspond to sub-storm period with time zone from 07:27 UT, 20/11/2001 until 03:00 UT, 21/11/2003.

Magnetic clouds are a possible manifestation of a Coronal Mass Ejection (CME) and they represent on third of ejectra observed by satellites. Magnetic cloud behaves like a magnetosphere moving through the solar wind. In Fig. 7[a-f] we showed the estimation of q-triplet for Bz magnetic field component. As we can observe the theoretical estimation is faithful with high precision on the whole part of the experimental function $f(a)$ and $D_q$. Additional, the q-triplet values were found to satisfy the relation $q_{rel} < 1 < q_{stat} < q_{rel}$ for the data set from solar wind magnetic cloud dynamics (see table 1).

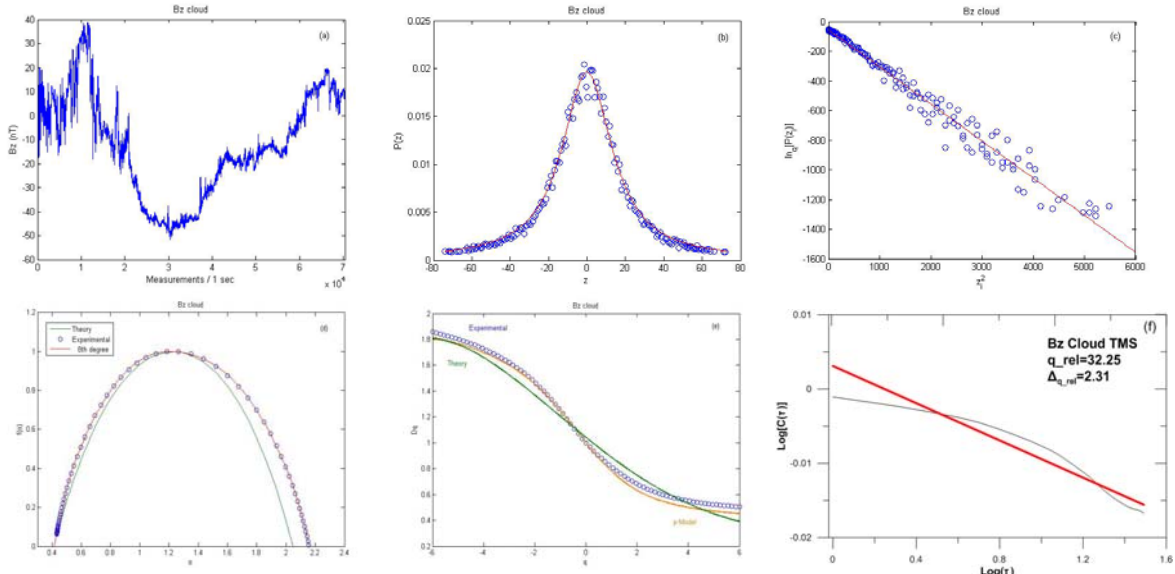

***Figure 7:*** *(a) Time series of Bz cloud (b) PDF $P(z_i)$ vs. $z_i$ q Gaussian function that fits $P(z_i)$ for the Bz cloud (c) Linear Correlation between $ln_q P(z_i)$ and $(z_i)^2$ where $q_{stat} = 2.02 \pm 0.04$ for the Bz cloud. d) Multifractal spectrum of Bz cloud time series with solid line a eighth degree polynomial. We calculate the $q_{sen} = 0.484 \pm 0.009$. (e) D(q) vs. q of the Bz cloud time series (f) Log – log plot of the self-correlation coefficient $C(\tau)$ vs. time delay $\tau$ for the Bz cloud time series. We obtain the best fit with $q_{rel} = 32.25 \pm 2.31$.*

### 4.9 Solar Dynamics

In this sub-section we present the q-triplet of the sunspot and solar flares complex systems by using data of Wolf number and daily Flare Index. Especially, we use the Wolf number, known as the international sunspot number measures the number of sunspots and group of sunspots on the surface of the sun computed by the formula: (10)*R=k\*(10g+s)* where: *s* is the number of individual spots, *g* is the number of sunspot groups and *k* is a factor that varies with location known as the observatory factor. We analyse a period of 184 years. Moreover we analyse the daily Flare Index of the solar activity that was determined using the final grouped solar flares obtained by NGDC (National Geophysical Data Center). It is calculated for each flare using the formula: $Q = (i*t)$, where "*i*" is the importance coefficient of the flare and "*t*" is the duration of the flare in minutes. To obtain final daily values, the daily sums of the index for the total surface are divided by the total time of observation of that day. The data covers time period from 1/1/1996 to 31/12/2007.

We obtain similar results from the previous physical systems in q-triplet Tsallis statistics. In additional, we clearly observe non-Gaussian statistics for both cases but the non-Gaussianity of solar flares was found much stronger than the sunspot index. The q-triplet values satisfy the relation $q_{rel} < 1 < q_{stat} < q_{rel}$ for the sunspot and solar flares timeseries (see table 1).

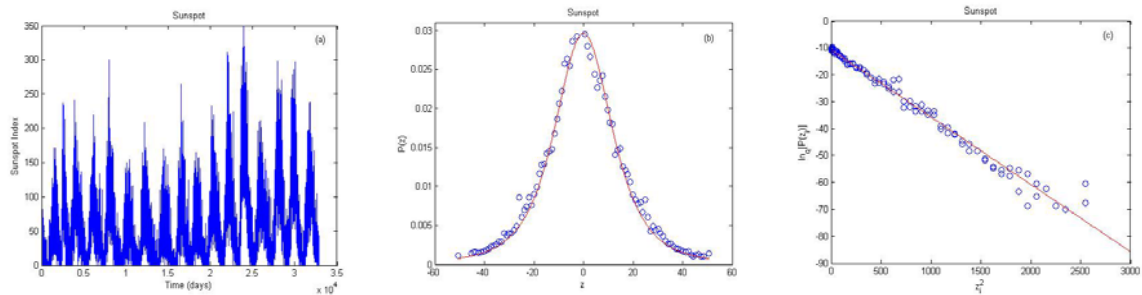

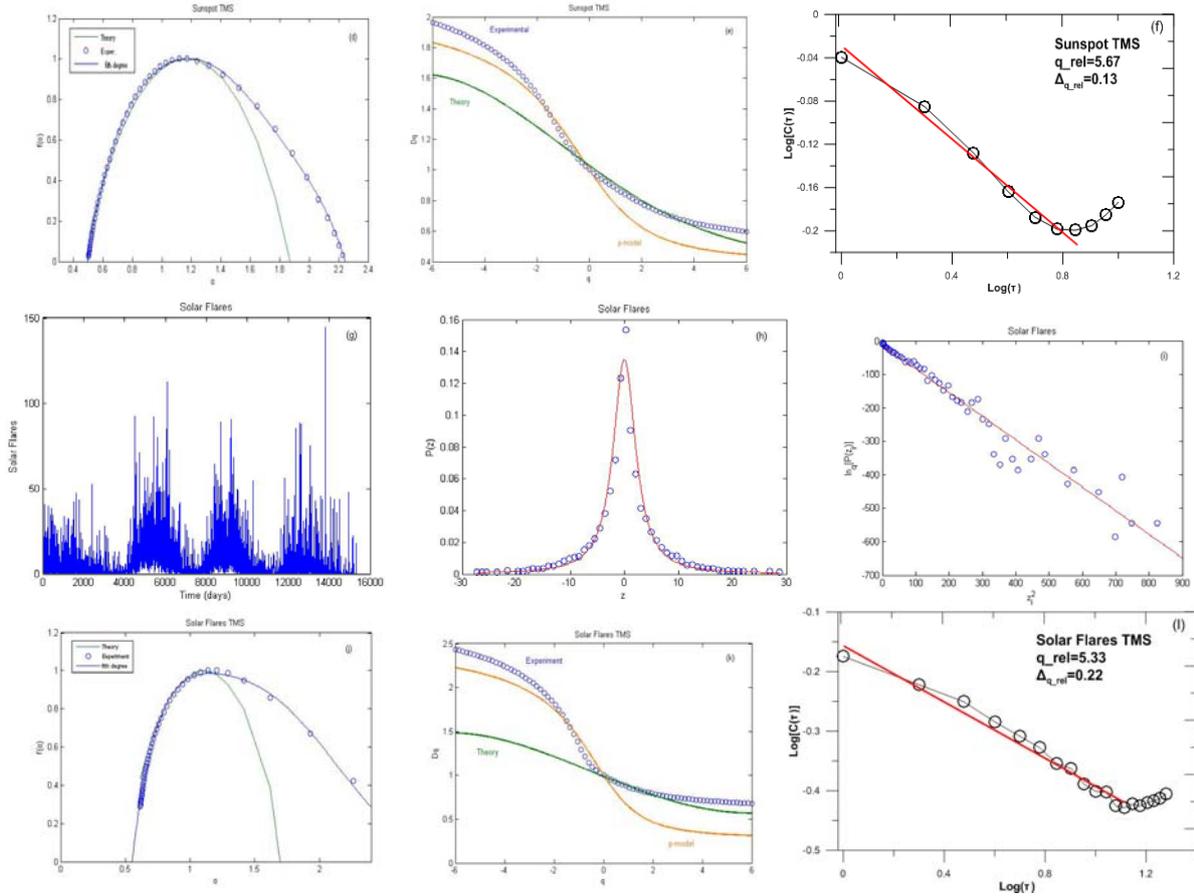

***Figure 8***: *(a) Time series of Sunspot Index concerning the period of 184 years (b) PDF $P(z_i)$ vs. $z_i$ q Guassian function that fits $P(z_i)$ for the Sunspot Index (c) Linear Correlation between $ln_q P(z_i)$ and $(z_i)^2$ where $q_{stat}$ = 1.53 ± 0.04 for the Sunspot Index (d) Multifractal spectrum of Sunspot Index time series with solid line a sixth degree polynomial. We calculate the $q_{sen}$ = 0.368 ± 0.005 (e) D(q) vs. q of the Sunspot Index time series (f) Log – log plot of the self-correlation coefficient $C(\tau)$ vs. time delay $\tau$ for the Sunspot Index time series. We obtain the best fit with $q_{rel}$ = 5.67 ± 0.13 (g) Time series of Solar Flares concerning the period of 184 years (h) PDF $P(z_i)$ vs. $z_i$ q Guassian function that fits $P(z_i)$ for the Solar Flares (i) Linear Correlation between $ln_q P(z_i)$ and $(z_i)^2$ where $q_{stat}$ = 1.90 ± 0.05 for the Solar Flares. (j) Multifractal spectrum of Solar Flares time series with solid line a sixth degree polynomial. We calculate the $q_{sen}$ = 0.308 ± 0.005. (k) D(q) vs. q of the Solar Flares time series (l) Log – log plot of the self-correlation coefficient $C(\tau)$ vs. time delay $\tau$ for the Solar Flares time series. We obtain the best fit with $q_{rel}$ = 5.33 ± 0.22.*

### 4.10 Solar Charged Particles Dynamics

In the following we present significant verification of theoretical prediction of Tsallis theory by study the q-triplet of energetic particle acceleration. We analyze energetic particles from spacecraft ACE – experiment EPAM and time zone 1997 day 226 to 2006 day 178 and protons (0.5 – 4) MeV with period 20/6/1986 – 31/5/2006, spacecraft GOES, hourly averaged data.

In Fig. 9[a-l] we showed the estimation of q-triplet for solar charged particles ($p^+$, $e^-$). We can observe the theoretical estimation is faithful with medium precision on the whole part of the experimental function $f(a)$ and $D_q$. Additional, the q-triplet values were found to satisfy the relation $q_{rel} < 1 < q_{stat} < q_{rel}$ for the data set from solar charged particles dynamics (see table 1).

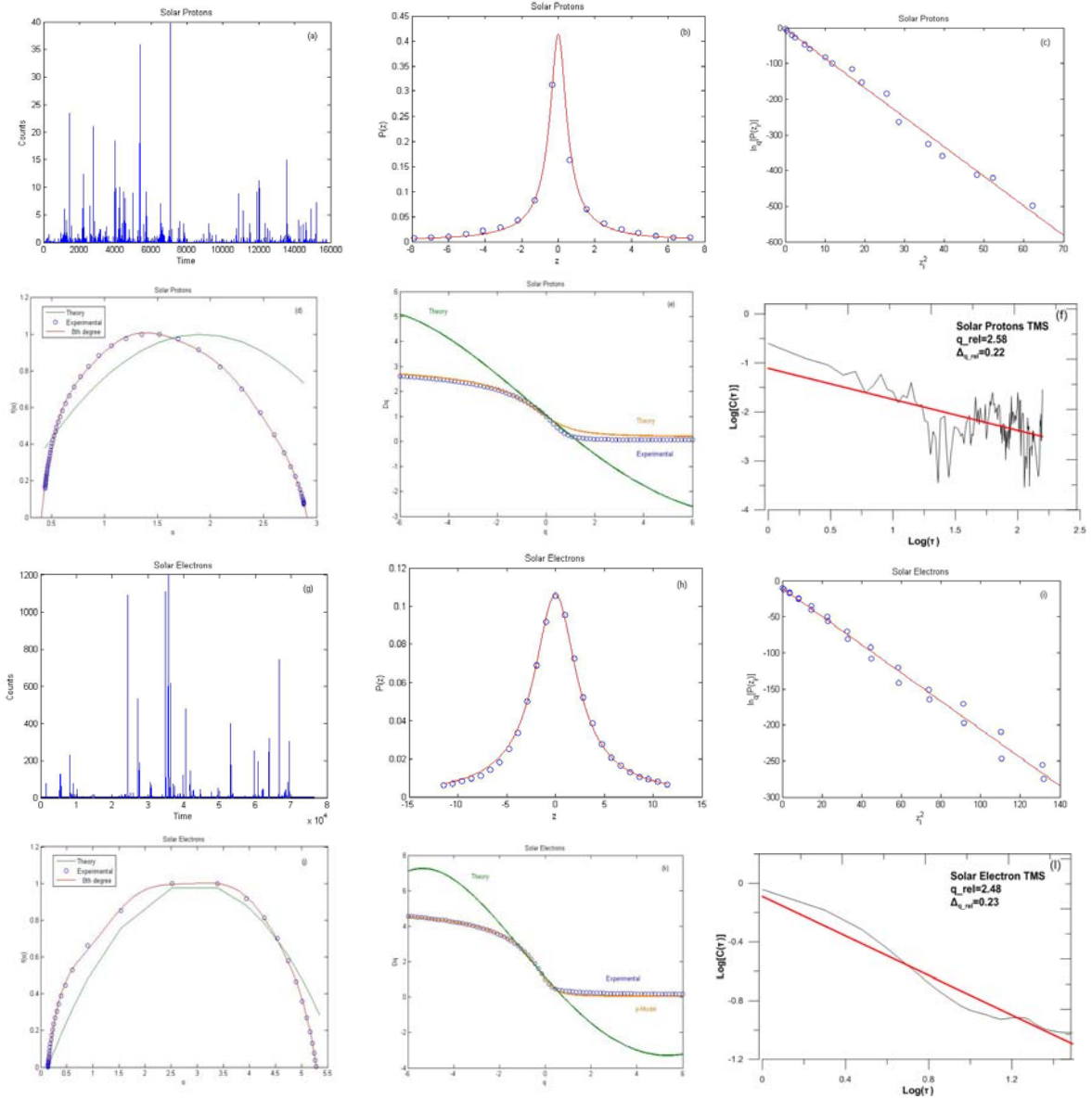

*Figure 9: (a)* Time series of Solar proton *(b)* PDF $P(z_i)$ vs. $z_i$ q Guassian function that fits $P(z_i)$ for the Solar proton data *(c)* Linear Correlation between $ln_q P(z_i)$ and $(z_i)^2$ where $q_{stat} = 2.31 \pm 0.13$ for the Solar proton *(d)* Multifractal spectrum of Solar proton time series with solid line a eighth degree polynomial. We calculate the $q_{sen} = 0.951 \pm 0.017$. *(e)* $D(q)$ vs. q of the Solar proton time series *(f)* Log – log plot of the self-correlation coefficient $C(\tau)$ vs. time delay $\tau$ for the Solar proton time series. We obtain the best fit with $q_{rel} = 2.58 \pm 0.22$ *(g)* Time series of Solar electrons *(h)* PDF $P(z_i)$ vs. $z_i$ q Guassian function that fits $P(z_i)$ for the Solar electrons *(i)* Linear Correlation between $ln_q P(z_i)$ and $(z_i)^2$ where $q_{stat} = 2.13 \pm 0.06$ for the Solar electrons *(j)* Multifractal spectrum of Solar electrons time series with solid line a eighth degree polynomial. We calculate the $q_{sen} = 0.860 \pm 0.009$. *(k)* $D(q)$ vs. q of the Solar electrons time series *(l)* Log – log plot of the self-correlation coefficient $C(\tau)$ vs. time delay $\tau$ for the Solar electrons time series. We obtain the best fit with $q_{rel} = 2.48 \pm 0.23$.

### 4.11 Cosmic Stars Dynamics

In the following we study the q-triplet Tsallis statistics for cosmic star brightness. For this we used a set of measurements of the light curve (time variation of the intensity) of the variable white dwarf star PG1159-035 during March 1989. It was recorded by the Whole Earth Telescope (a coordinated group of telescopes distributed around the earth that permits the continuous observation of an astronomical object) and submitted by James Dixson and Don Winget of the Department of Astronomy and the McDonald Observatory of the University of Texas at Austin. The telescope is described in an article in The Astrophysical Journal (361), p.

309-317 (1990), and the measurements on PG1159-035 will be described in an article scheduled for the September 1 issue of the Astrophysical Journal. The observations were made of PG1159-035 and a non-variable comparison star. A polynomial was fit to the light curve of the comparison star, and then this polynomial was used to normalize the PG1159-035 signal to remove changes due to varying extinction (light absorption) and differing telescope properties.

In Fig.11 we showed the results from q-triplet analysis for cosmic star. As we can observe the theoretical estimation is faithful with high precision on the whole part of the experimental function $f(a)$ and $D_q$. Moreover the q-triplet values were found to satisfy the relation $q_{rel} < 1 < q_{stat} < q_{rel}$ for the data set cosmic star (see table 1).

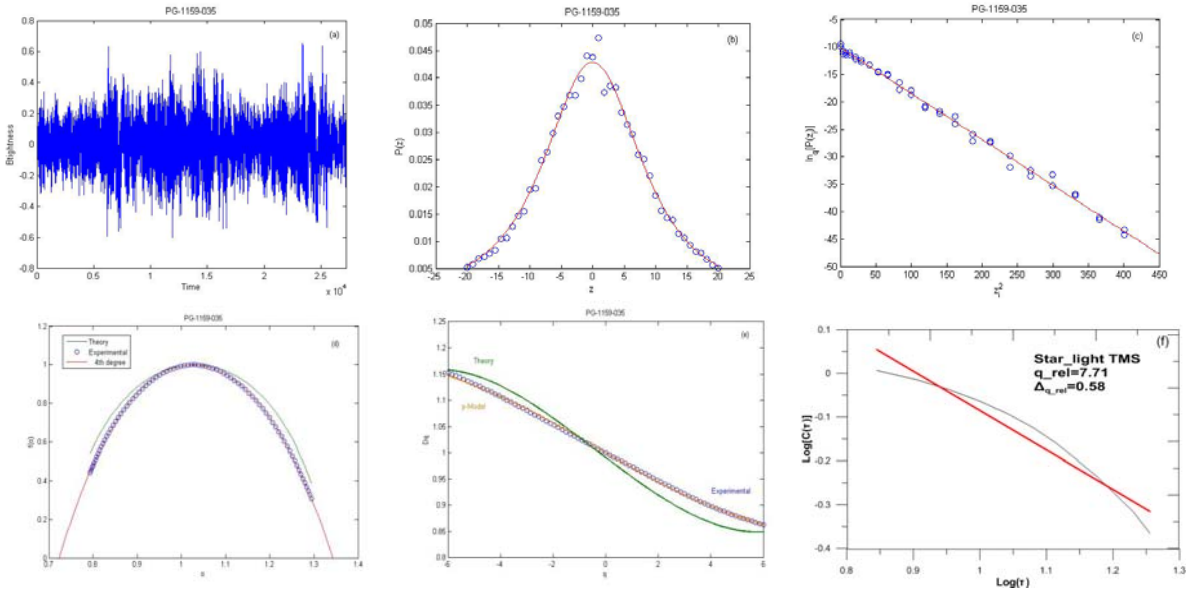

***Figure 10:*** *(a) Time series of cosmic star PG-1159-035 (b) PDF $P(z_i)$ vs. $z_i$ q Guassian function that fits $P(z_i)$ for the cosmic star PG-1159-035 (c) Linear Correlation between $\ln_q P(z_i)$ and $(z_i)^2$ where $q_{stat} = 1.64 \pm 0.03$ for the cosmic star PG-1159-035 (d) Multifractal spectrum of the cosmic star PG-1159-035 time series with solid line a fourth degree polynomial. We calculate the $q_{sen} = -0.568 \pm 0.042$ (e) D(q) vs. q of the cosmic star PG-1159-035 time series (f) Log – log plot of the self-correlation coefficient C(τ) vs. time delay τ for the cosmic star PG-1159-035 time series. We obtain the best fit with $q_{rel} = 7.71 \pm 0.58$.*

**4.12 Cosmic Rays Dynamics**

In this sub-section we study the q-triplet for the cosmic ray (carbon) data set. For this we used the data from the Cosmic Ray Isotope Spectrometer (CRIS) on the Advanced Composition Explorer (ACE) spacecraft and especially the carbon element (56-74 Mev) in hourly time period and time zone duration from 2000 – 2011.The cosmic rays data set is presented in Fig.11a, while the q-triplet is presented in Fig.11[b,f]. This result reveals clearly non-Gaussian statistics for the cosmic rays data. The theoretical estimation is faithful with high precision on the whole part of the experimental function $f(a)$ and $D_q$. Similar, the q-triplet values were found to satisfy the relation $q_{rel} < 1 < q_{stat} < q_{rel}$ (see table 1).

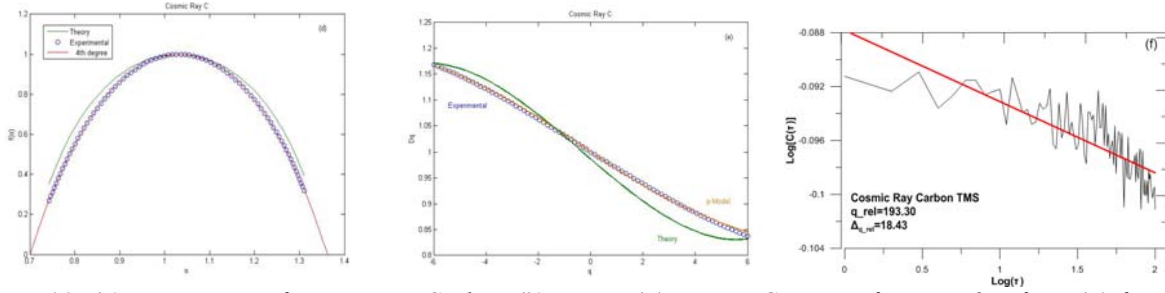

***Figure 11:*** *(a) Time series of cosmic ray Carbon (b) PDF $P(z_i)$ vs. $z_i$ q Guassian function that fits $P(z_i)$ for the cosmic ray Carbon (c) Linear Correlation between $\ln_q P(z_i)$ and $(z_i)^2$ where $q_{stat} = 1.44 \pm 0.05$ for the cosmic ray Carbon (d) Multifractal spectrum of the cosmic ray Carbon time series with solid line a fourth degree polynomial. We calculate the $q_{sen} = -0.441 \pm 0.013$. (e) $D(q)$ vs. q of the cosmic ray Carbon time series (f) Log – log plot of the self-correlation coefficient $C(\tau)$ vs. time delay $\tau$ for the cosmic ray Carbon time series. We obtain the best fit with $q_{rel} = 193.30 \pm 18.43$.*

**Table 1**

| System | q_sen | q_stat | q_rel (C(τ)) |
|---|---|---|---|
| Cardiac (hrv) | -0.772±0.068 | 1.26±0.10 | 3.32±0.44 |
| Brain (seizure) | -0.451±0.036 | 1.63±0.14 | 10.43±0.95 |
| Seismic (Interevent) | 0.370±0.005 | 2.28±0.12 | 2.89±0.26 |
| Seismic (Magnitude) | -0.183±0.027 | 1.77±0.09 | 7.53±0.54 |
| Atmosphere (Temperature) | 0.407±0.013 | 1.89±0.08 | 8.87±0.62 |
| Atmosphere (Rainfall) | 0.444±0.007 | 2.21±0.06 | 6.04±0.47 |
| Magnetosphere (Bz storm) | 0.341±0.033 | 2.05±0.04 | 6.45±0.35 |
| Magnetosphere (Vx storm) | 0.516±0.010 | 1.98±0.06 | 2.30±0.12 |
| Magnetosphere (Ey storm) | 0.685±0.016 | 2.49±0.07 | 3.95±0.34 |
| Magnetosphere (Electrons storm) | -0.838±0.003 | 2.15±0.07 | 13.19±0.95 |
| Magnetosphere (Protons storm) | 0.533±0.017 | 2.49±0.05 | 9.13±0.67 |
| Solar Wind (Bz cloud) | 0.484±0.009 | 2.02±0.04 | 32.25±2.31 |
| Solar (Sunspot Index) | 0.368±0.005 | 1.53 ± 0.04 | 5.672±0.127 |
| Solar (Flares Index) | 0.308±0.005 | 1.870±0.005 | 5.33±0.22 |
| Solar (Protons) | 0.951±0.017 | 2.31±o.13 | 2.48±0.23 |
| Solar (Electrons) | 0.860±0.009 | 2.13±0.06 | 2.58±0.22 |
| Cosmic Stars (Brigtness) | -0.568±0.042 | 1.64±0.03 | 7.71±0.58 |
| Cosmic Ray (C) | -0.441±0.013 | 1.44±0.05 | 193.30±18.43 |

***Table 1:*** *Summarize parameter values of dynamics from different kinds of physical systems: The q-triplet (q_sen, q_stat, q_rel of Tsalli's.*

## 5. Summary and Discussion

In this study we have presented theoretical and observational indications for the new status of non-equilibrium space plasma theory. Especially, the Tsallis non-extensive $q$ – statistics have been verified in every case: the magnetospheric plasma, the solar wind, the solar plasma (convection zone, solar corona) for cosmic stars and cosmic rays. As the Tsallis $q$ – statistics reveals long range correlations and strong self-organization process we presented also the highlights of non-equilibrium statistical theory of random fields putting emphasis to the theory of $n$ – point correlations, which are responsible for the complex character of the non-equilibrium dynamics. In the following we present further discussion of the space plasma complex processes in relation with the general transformation of modern scientific concepts included in the complexity theory.

For all cases the statistics of experimental space plasma signal was found to obey the non-extensive q-statistics of Tsallis with high q-values ($q_{stat} \simeq 1.50 - 2.50$). Furthermore, the q-statistics for the magnetospheric system was estimated to obtain q-value ($q_{stat} \simeq 2.00$) for MHD signals (magnetic field bulk plasma flow) and $q_{stat} \simeq 2.50$ for electric field and energetic particles. For solar flares activity and solar energetic particles the q-values were also found to be higher than two ($q_{stat} \simeq 1.90 - 2.30$). For the solar convention zone (sunspot index) the q-value was found to be $q_{stat} = 1.53$, while for the cosmic rays (carbon experimental data) the q-value was found to be $q_{stat} = 1.44$. The above q-statistics values reveal clearly that space plasma dynamics obeys Tsallis non-extensive statistics in every case, from planetary magnetospheres to solar plasma and solar corona as well as to cosmic stars and cosmic rays.

These results are in wonderful agreement and harmony with the theoretical framework presented in previous sections of this study. In fact, the presence of q-statistics indicates clearly the necessity of fractal generalization of dynamics for the theoretical interpretation of non-equilibrium space plasma dynamics in accordance with many theoretical studies that are cited here. Finally, we believe that the Tsallis q-statistical theory coupled with the non-equilibrium fractal generalization of dynamics indicates a novel road for the space plasmas science. Moreover, this point of research view highlights the space plasmas system as one of the most significant case for the application of new theoretical concepts introduced by the complexity science during the last two or three decades.

## Acknowledgements

We thank the ACE MAG instrument team and the ACE Science Center for providing the ACE data [72]. Moreover, we thank S. Kokubun and T. Mukai for the high resolution Geotail/MGF magnetic field and Geotail/LEP plasma data, respectively.